\documentclass[8pt]{article}
\usepackage[margin=0.5in]{geometry}
\usepackage{amsmath,amsthm,amssymb}
\usepackage{stmaryrd}
\usepackage{graphicx}
\usepackage{hyperref}
\usepackage{booktabs}
\usepackage{mathpazo}
\usepackage{tikz}
\usepackage{pgfplots}
\usepackage{pgf,tikz}
\pgfplotsset{compat=1.14}
\usepackage{enumitem}
\usepackage{lipsum}
\usepackage{braket}
\usepackage{algorithm}
\usepackage{algorithmic}
\graphicspath{{data/}}

\usepackage{hyperref}
\hypersetup{
    colorlinks=true,
    linkcolor=blue,
    filecolor=magenta,      
    urlcolor=cyan,
    pdftitle={Overleaf Example},
    pdfpagemode=FullScreen,
}

\theoremstyle{definition}

\newcommand{\code}{\mathcal{C}}
\newcommand{\proj}{\Pi}
\newcommand{\ecset}{\mathcal{E}_\textrm{C}}
\newcommand{\edset}{\mathcal{E}_\textrm{D}}
\newcommand{\trace}{\mathrm{Tr}}

\title{Transversal Gates in Nonadditive Quantum Codes}
\author{Chao Zhang,\   Zipeng Wu,\  Shilin Huang\\ 
Department of Physics, The Hong Kong University of Science and Technology, \\
Clear Water Bay, Kowloon, Hong Kong\\\small\\
Bei Zeng\\ Department of Physics, The University of Texas at Dallas,\\ Richardson, Texas 75080, USA}
\date{\today}

\begin{document}

\maketitle

\begin{abstract}
Transversal gates play a crucial role in suppressing error propagation in fault‑tolerant quantum computation, yet they are intrinsically constrained: any nontrivial code encoding a single logical qubit admits only a finite subgroup of $\mathrm{SU}(2)$ as its transversal operations. We introduce a systematic framework for searching codes with specified transversal groups by parametrizing their logical subspaces on the Stiefel manifold and minimizing a composite loss that enforces both the Knill–Laflamme conditions and a target transversal‑group structure.  Applying this method, we uncover a new $((6,2,3))$ code admitting a transversal $Z\bigl(\tfrac{2\pi}{5}\bigr)$ gate (transversal group $\mathrm{C}_{10}$), the smallest known distance $3$ code supporting non‑Clifford transversal gates, as well as several new $((7,2,3))$ codes realizing the binary icosahedral group $2I$. We further propose the \emph{Subset‑Sum–Linear‑Programming} (SS‑LP) construction for codes with transversal \emph{diagonal} gates,  which dramatically shrinks the search space by reducing to integer partitions subject to linear constraints. In a more constrained form, the method also applies directly to the binary-dihedral groups $\mathrm{BD}_{2m}$.  Specializing to $n=7$, the SS‑LP method yields codes for all $\mathrm{BD}_{2m}$ with $2m\le 36$,  including the first $((7,2,3))$ examples supporting transversal $T$ gate ($\mathrm{BD}_{16}$) and $\sqrt{T}$ gate ($\mathrm{BD}_{32}$), improving on the previous smallest examples $((11,2,3))$ and $((19,2,3))$. Extending the SS-LP approach to $((8,2,3))$, we construct new codes for $2m>36$, including one supporting a transversal $T^{1/4}$ gate ($\mathrm{BD}_{64}$).  These results reveal a far richer landscape of nonadditive codes than previously recognized and underscore a deeper connection between quantum error correction and the algebraic constraints on transversal gate groups.  
\end{abstract}

\section{Introduction}

Quantum error correction (QEC) is indispensable for fault‑tolerant quantum
computation~\cite{calderbank1998quantum}. It shields fragile quantum states from noise and decoherence.
A key ingredient is the use of \emph{transversal} logical gates, in which
each physical qubit is acted on by an independent local unitary so that a
single‑qubit fault cannot spread across the code block. Although the
Knill--Laflamme (KL) error‑correction conditions~\cite{knill1997theory} and the design of specific
transversal gates might seem unrelated, foundational no‑go results already
hint at a connection: any non‑trivial code is limited to a \emph{finite} set
of transversal logical operations and can never realize a universal gate set
by transversal means alone \cite{eastin2009restrictions,zeng2011transversality}.

In this work, we concentrate on codes with parameters \(\bigl((N,2,d>2)\bigr)\),
which use \(N\) physical qubits to encode a single logical qubit. In this scenario, the only possible transversal gate groups are the finite subgroups of SU(2). These include the cyclic groups \(\mathrm{C}_{2m}\), which consist of phase rotations; the binary dihedral groups \(\mathrm{BD}_{2m}\), which add reflections generated by Pauli operators; and the three exceptional subgroups related to the symmetry groups of Platonic solids: the binary tetrahedral group (\(2T\)), the binary octahedral group (\(2O\)), and the binary icosahedral group (\(2I\)). Each finite subgroup admits a distinct set of logical gates that can potentially be realized transversally in a distance-\(d\) code.

We place particular emphasis on \emph{nonadditive} constructions. Nonadditive codes~\cite{rains1997nonadditive,roychowdhury1997structure,grassl1997note,arvind2002nonstabilizer,aggarwal2008boolean,ni2015non,webster2022xp} are
still much less charted than stabilizer codes, yet they can support
transversal gate sets that lie far outside the stabilizer paradigm, such as the binary icosahedral group $2I$ ~\cite{kubischta2023family}. 
In fact,  the set of diagonal gates that can implement a single or two-qubit transversal
logical gate for qubit stabilizer codes has been classified~\cite{zeng2011transversality,anderson2016classification}. 
In particular, individual physical gates on the underlying
qubits that compose the code are restricted to have entries of the form $e^{i\pi c/2^k}$
along their diagonal. Consequently, all diagonal logical gates that can be implemented transversally by individual
physical diagonal gates must belong to the Clifford hierarchy~\cite{anderson2016classification}. Furthermore,
the smallest stabilizer code supporting the transversal $T$ gate is the $((15,2,3))$ Reed-Muller code~\cite{koutsioumpas2024quantum},
while there exists $((11,2,3))$ permutation invariant code supporting transversal $T$ gate~\cite{kubischta2024permutationinvariantquantumcodestransversal}.

Familiar representatives of nonadditive codes are the codeword--stabilised (CWS) codes
\cite{cross2009codeword,yu2007graphical,yu2008nonadditive,grassl2008non,chuang2009codeword} and the permutation invariant codes~\cite{ouyang2014permutation,ouyang2021permutation} first proposed by Pollatsek and
Ruskai~\cite{pollatsek2004permutationally}.  
Recent work introduced a local unitary (LU)‑invariant norm 
that parametrises families of nonadditive codes: 
\(
\lambda^{\ast}\;=\;\sqrt{\sum_{i<j}|\lambda_{ij}|^2}.
\)
Here, \(\lambda_{ij}\) ($i<j$) are the off‑diagonal coefficients
in the KL condition $\proj E_i^\dag E_j \proj = \lambda_{ij} P$~\cite{du2024characterizing}, which intuitively quantify the correlations of errors within the code subspace. One can show that 
two codes with different values of $\lambda^\ast$ cannot be equivalent up to LU transformations. 
Because stabilizer codes force each $\lambda_{ij}$ to take discrete values, one finds that $\bigl(\lambda^{\ast}\bigr)^2$ must be an integer: in the $((6,2,3))$ setting, the unique stabilizer solution has $(\lambda^{\ast})^2 = 1$, corresponding to degenerate codes. While in the $((7,2,3))$ case, stabilizer constructions realise only
$
(\lambda^{\ast})^2 \in \{0,1,2,3,5\}
$~\cite{cross2025small},
recovering the nondegenerate Steane code at $\lambda^\ast=0$.
By contrast, nonadditive codes fill out continuous intervals, which hints at a rich continuum of
examples of different nonadditive codes. For \(((6,2,3))\) codes, one finds
\(\lambda^\ast\in[\sqrt{0.6},\,1]\); every value in this interval is realised
by a continuous family of distinct codes. For \(((7,2,3))\) codes, the range
widens to \(\lambda^\ast\in[0,\sqrt{7}]\), interpolating smoothly from the
non‑degenerate Steane code \((\lambda^\ast=0)\) to the
permutation‑invariant code \((\lambda^\ast=\sqrt{7})\) by Pollatsek and
Ruskai~\cite{pollatsek2004permutationally} (which also coincides with the \(((7,2,3))\) code realising
the transversal group \(2I\)~\cite{kubischta2023family}).
These results go well beyond the CWS and permutation‑invariant
constructions and underscore the existence of vast, previously uncatalogued
regions of nonadditive codes--and therefore of new candidate transversal
gate groups--waiting to be explored.

We present a systematic methodology to search for potential quantum codes that simultaneously satisfy QEC requirements and admit desired transversal gate groups. In our approach, we represent each candidate code subspace as a point on the Stiefel manifold of orthonormal two‑frames in the \(2^{n}\)-dimensional Hilbert space and minimize a loss function consisting of two terms: one term penalizes violation of the KL conditions for a specified error set (e.g., all Pauli errors of weight \(\le 2\)), while the other penalises deviations from the target transversal group.  
Updates based on polar or exponential maps preserve both the orthonormality of the code vectors and the unitarity of the local operators, allowing us to balance pure QEC performance against the desired transversal structure. Applying this method, we uncover several novel phenomena. We find a new \(\bigl((6,2,3)\bigr)\) code whose transversal group is \(\mathrm{C}_{10}\), representing the smallest known example of distance $3$ codes supporting a non-Clifford gate of order 10. We identify two previously unknown \(\bigl((7,2,3)\bigr)\) codes that realize the binary icosahedral group \(2I\) at signature norms \(\lambda^*=0\) and \(\lambda^*=\sqrt{3/4}\). We also locate two-dimensional subspaces of seven qubits that admit transversal \(2I\) operations yet fail the KL test, illustrating subtle obstructions to full error correction.

We further develop the \emph{Subset-Sum–Linear-Programming} (SS-LP) construction for codes with transversal \emph{diagonal} gates. By recasting the search as integer-partition problems subject to simple linear constraints, SS-LP dramatically shrinks the optimization domain. In a more constrained form, the method also applies directly to the binary-dihedral groups $\mathrm{BD}_{2m}$. For $n=7$, the SS-LP method yields binary-dihedral groups with $2m \leq 36$, including transversal $T$ ($\mathrm{BD}_{16}$) and $\sqrt{T}$ ($\mathrm{BD}_{32}$), improving upon the previously known smallest codes implementing transversal $T$ and $\sqrt{T}$, which were $((11,2,3))$ and $((19,2,3))$ respectively~\cite{kubischta2024permutationinvariantquantumcodestransversal}. The SS-LP method extends to larger blocks, as illustrated by $((8,2,3))$ codes with transversal group $\mathrm{BD}_{2m}$ for $2m > 36$, including one supporting transversal $T^{1/4}$ ($\mathrm{BD}_{64}$).

These findings expose a subtler relationship between transversal gate groups
and the KL conditions than previously recognised and greatly expand the
catalogue of nonadditive codes worth exploring in future fault‑tolerant
architectures.

\section{Preliminaries}\label{pre}

In this section, we briefly review standard notions and known results on quantum error-correcting codes (QECCs) that will be used throughout the paper.

\subsection{Quantum Error-Correcting Codes}

\paragraph{Knill--Laflamme Conditions.}
Let \(\ecset = \{E_i\}\) denote a specified set of error operators (e.g., Kraus operators of a noisy quantum channel). A quantum error-correcting code (QECC) is defined by a projector \(\proj\) onto a subspace \(\code\) --- called the \emph{code subspace} --- such that the \emph{Knill--Laflamme (KL) condition} holds:
\begin{equation}
\proj E_i^\dagger E_j \proj = \lambda_{ij} \proj,
\quad \forall\, E_i, E_j \in \ecset,
\label{eq:knill-laflamme}
\end{equation}
where \(\lambda_{ij}\) are complex scalars.

This condition guarantees that any error from \(\ecset\) can be perfectly corrected, provided it acts within the code subspace \(\code\). Moreover, the entire linear span of \(\ecset\), denoted by \(\mathrm{span}\;\ecset \subset \mathbb{C}^{2^n \times 2^n}\), consists of correctable errors.

If \(\code\) has dimension \(K\), then \(\proj\) is a rank-\(K\) projector on the \(2^n\)-dimensional Hilbert space, and can be expressed as
\[
\proj = \sum_{i=0}^{K-1} \ket{\psi_i}\bra{\psi_i},
\]
where \(\{\ket{\psi_i}\}\) is an orthonormal basis for \(\code\). In terms of this basis, the KL condition~\eqref{eq:knill-laflamme} is equivalent to:
\begin{equation}
\bra{\psi_k} E_i^\dagger E_j \ket{\psi_\ell} = \lambda_{ij} \delta_{k\ell},
\quad \forall\, E_i, E_j \in \ecset,\; 0 \le k,\ell < K.
\label{eq:knill-laflamme2}
\end{equation}

The matrix \((\lambda_{ij})\) characterizes how pairs of errors \(E_i, E_j\) act on the code subspace. A QECC is called \emph{nondegenerate} if \((\lambda_{ij})\) is a full-rank (i.e., nonsingular) Hermitian matrix, meaning the errors are distinguishable when restricted to \(\code\). Conversely, the code is \emph{degenerate} if \((\lambda_{ij})\) is singular, implying that some errors act identically on \(\code\) and are thus indistinguishable. In the extreme case of \emph{complete degeneracy}, known as a \emph{decoherence-free subspace} (DFS), the matrix \((\lambda_{ij})\) may have rank \(1\)~\cite{lidar1998decoherence,lidar2003decoherence}.

Furthermore, if two errors \(E_i, E_j \in \ecset\) are correctable by the code \(\code\), then their product \(E_i^\dagger E_j\) is \emph{detectable}. That is, the QECC also functions as a quantum \emph{error-detecting code}. An error operator \(E \in \mathbb{C}^{2^n \times 2^n}\) is detectable if and only if there exists a scalar \(\lambda_E \in \mathbb{C}\) such that:
\begin{equation}
\proj E \proj = \lambda_E \proj,
\label{eq:knill-laflamme3}
\end{equation}
or equivalently,
\begin{equation}
\bra{\psi_i} E \ket{\psi_j} = \lambda_E \delta_{ij}.
\label{eq:knill-laflamme4}
\end{equation}

Equations~\eqref{eq:knill-laflamme3} and~\eqref{eq:knill-laflamme4} are the KL conditions for detectable errors, and they serve as the basis for QECC search algorithms in this work. One can define a corresponding \emph{detectable error set} \(\edset\), consisting of a basis of detectable errors. It is always true that
\[
E_i^\dagger E_j \in \mathrm{span}\;\edset \quad \forall E_i, E_j \in \ecset.
\]
However, it is also possible that:
\[
\mathrm{span}\;\edset \supsetneq \mathrm{span} \left\{ E_i^\dagger E_j : E_i, E_j \in \ecset \right\},
\]
i.e., the space of detectable errors may be strictly larger than the span of error products.

\paragraph{Pauli Errors and Weight.} 
Quantum errors are oftentimes modeled using tensor products of single-qubit Pauli operators:
\[
I = \begin{bmatrix}1 & 0\\ 0 & 1\end{bmatrix},\quad
X = \begin{bmatrix}0 & 1\\ 1 & 0\end{bmatrix},\quad
Y = \begin{bmatrix}0 & -i\\ i & 0\end{bmatrix},\quad
Z = \begin{bmatrix}1 & 0\\ 0 & -1\end{bmatrix}.
\]
The set \(\{I, X, Y, Z\}^{\otimes n}\) consists of all \(n\)-fold tensor products of these operators and serves as a convenient basis for describing errors on \(n\)-qubit systems. We refer to these as \emph{Pauli error operators}. 
A Pauli error operator \(P = P_1  \otimes \cdots \otimes P_n \in \{I, X, Y, Z\}^{\otimes n}\), is said to have \emph{weight} \(\mathrm{wt}(P)\), defined as the number of positions \(i\) for which \(P_i \ne I\). This counts the number of qubits on which the operator acts nontrivially.

For a general $n$-qubit operator \(E \in \mathbb{C}^{2^n \times 2^n}\), \(E\) can be uniquely written as a linear combination \(E = \sum_{P} \alpha_P P\),
where the sum is over all \(n\)-qubit Pauli error operators \(P \in \{I, X, Y, Z\}^{\otimes n}\), and \(\alpha_P \in \mathbb{C}\). The \emph{weight} of \(E\), denoted \(\mathrm{wt}(E)\), is defined as:
\[
\mathrm{wt}(E) = \max \left\{ \mathrm{wt}(P) : \alpha_P \ne 0 \right\}.
\]

\paragraph{Code Distance.} 
The \emph{distance} \(d\) of a QECC quantifies its ability to detect and correct errors. It is defined as the minimum weight of an error operator \(E \in \{I, X, Y, Z\}^{\otimes n}\) for which the Knill-Laflamme condition~\eqref{eq:knill-laflamme3} is violated:
\[
d = \min \left\{ \mathrm{wt}(E) : \proj E \proj \not\propto \proj \right\}.
\]
Equivalently, \(d\) is the minimum weight of a nontrivial operator acts nontrivially within \(\code\) (i.e., as a logical operator).
We use the notation $((n,K,d))$ to denote that an $n$-qubit code $\code$ of dimension $K$ has distance $d$.

A QECC of distance \(d\) can:
- Detect all Pauli errors of weight less than \(d\),
- Correct all such errors of weight up to \(\left\lfloor \frac{d - 1}{2} \right\rfloor\).
For example, the Steane code is a \(((7, 2, 3))\) code that can correct any single-qubit error and detect all errors of weight $\le 2$~\cite{steane1996error}.
Thus the sets $\ecset$ and $\edset$ can be chosen as
\begin{eqnarray*}
\ecset &=& \left\{ E \in \{I,X,Y,Z\}^{\otimes n}: 1\le \operatorname{wt}(E) \le \left\lfloor \frac{d-1}{2} \right\rfloor\right\}\\
\edset &=& \left\{ E \in \{I,X,Y,Z\}^{\otimes n}: 1\le \operatorname{wt}(E) \le d-1\right\}
\end{eqnarray*}

\paragraph{Non-Additive Codes.}  A QECC is called \emph{additive}, or equivalently a \emph{stabilizer code}, if it can be described as the joint \(+1\) eigenspace of an abelian subgroup of the \(n\)-qubit Pauli group~\cite{gottesman1997stabilizer}. In this case, if the code subspace has dimension \(K = 2^k\), the code encodes \(k\) logical qubits. In contrast, \emph{nonadditive} codes do not arise from abelian subgroups of the Pauli group and are not restricted to having dimension \(2^k\). Nevertheless, they still satisfy the KL conditions and can sometimes achieve higher encoding rates for fixed \(n\) and \(d\). Notable examples include certain codeword-stabilized (CWS) constructions, which have been shown to outperform comparable stabilizer codes in their encoding rate~\cite{rains1997nonadditive,     yu2007graphical,yu2008nonadditive,cross2009codeword}.

\subsection{Local equivalence, weight enumerators, and the signature vector}

Two QECCs \(\proj\) and \(\proj'\) are said to be \emph{locally equivalent} if there exists a local unitary
\[ U \;=\; U_1\otimes \cdots \otimes U_n \]
such that 
\[ \proj' \;=\; U\,\proj\,U^\dagger. \]
Local equivalence preserves the code parameters \(((n,K,d))\) but can change the way individual codewords appear.

A useful set of invariants for testing local equivalences is \emph{quantum weight enumerators}~\cite{shor1997quantum}, which measure how the projector \(\Pi\) (into a code space) overlaps with tensor products of Pauli operators \(\{I,X,Y,Z\}^{\otimes n}\). Specifically, one can define two polynomials $A(z) \;=\; \sum_{j=0}^n A_j \, z^j$,
$B(z) \;=\; \sum_{j=0}^n B_j \, z^j$,
where the coefficients
\begin{align}
A_j &= \frac{1}{K^2}\sum_{\operatorname{wt}(E)=j} \trace\bigl(E\,\proj\bigr)\,\trace\bigl(E^{\dagger}\,\proj\bigr),\\
B_j &= \frac{1}{K}\sum_{\operatorname{wt}(E)=j} \trace\bigl(E\,\proj\,E^{\dagger}\,\proj\bigr).
\end{align}
Here \(E\) runs through all weight-$j$ Pauli operators in $\{I,X,Y,Z\}^{\otimes n}$.
Related quantum weight enumerators, such as Rains' unitary and shadow enumerators~\cite{rains1998quantum,rains1999quantum,miller2024experimental}, provide additional invariants under local (or global) transformations.

From the KL condition $\proj E \proj = \lambda_E \proj$, we collect the scalars \[\lambda_E = \frac{1}{K} \trace\bigl(\proj E \proj\bigr) 
= \frac{1}{K} \trace\bigl(E \proj^2\bigr) = \frac{1}{K} \trace\bigl(E \proj\bigr),\quad \forall E \in \edset\]  into a vector
$\vec\lambda = (\lambda_E)_{E\in\edset}$.
We then define the \emph{signature norm} $\lambda^* = \|\vec\lambda\|_2$ as the Euclidean length of this "signature vector" \(\vec\lambda\)~\cite{du2024characterizing}. In case that $\proj$ is the projector of an $((n,K,d))$ code and \[\edset = \left\{E \in \{I,X,Y,Z\}^{\otimes n}: 1 \le \operatorname{wt}(E) \le d-1\right\},\]
one has
\[
\lambda^* =  \sqrt{\sum_{E \in \edset} |\lambda_E|^2} = \sqrt{\frac{1}{K^2}\sum_{E \in \edset} \trace\bigl(E\proj\bigr) \trace\bigl(E^\dagger\proj\bigr)} = \sqrt{\sum_{j=1}^{d-1} A_j}.
\]
Then \(\lambda^*(\Pi)\) as a function of the code  projector $\proj$ is a \emph{local unitary invariant} (LUI): if \(\lambda^*(\proj)\neq \lambda^*(\proj')\) for two codes \(\proj\) and \(\proj'\), then those codes must be locally inequivalent. 
Different codes may still share the same value of \(\lambda^*\) but be locally inequivalent, but \(\lambda^*\) remains a valuable tool to distinguish many inequivalent QECCs.

\subsection{Transversal Logical Gates}

For an $((n,2,d))$ quantum error-correcting code (QECC) $\code$, two orthonormal states \(\ket{0_L}\) and \(\ket{1_L}\) in $\code \subset \mathbb{C}^{2^n}$ represent the logical states \(\ket{0}\) and \(\ket{1}\). The physical implementation of a logical unitary gate \(U \in U(2)\) is denoted by \(\overline{U} \in U(2^n)\). While the choice of \(\overline{U}\) might not be unique, it must satisfy the condition:
\[
U_{ij} = \bra{i}U\ket{j} = \bra{i_L} \overline{U}\ket{j_L}, \quad i,j \in \{0,1\}.
\]

A logical gate \(U\) is said to be \emph{transversal} if it can be implemented as a tensor product of single-qubit unitary gates:
\begin{equation}\label{eq:trans-logical-def}
    \overline{U} = U_1 \otimes U_2 \otimes \cdots \otimes U_n,
\end{equation}
where each \(U_i \in U(2)\) acts on the \(i\)-th qubit.

Transversal logical gates are highly valued in fault-tolerant quantum computing because they prevent a single physical error from propagating to multiple qubits within the same code block. However, a fundamental no-go theorem states that no QECC can implement a universal gate set entirely through transversal operations~\cite{eastin2009restrictions,zeng2011transversality}. Specifically, for any $((n,2,d>1))$ code protecting a single logical qubit, the group of all transversal logical gates is a \emph{finite} subgroup of \(\mathrm{SU}(2)\) (up to an overall phase in \(\mathrm{U}(2)\)).

Since universality in \(\mathrm{SU}(2)\) requires an infinite and dense subgroup, the transversal logical gates of any QECC can only form one of the finite subgroups of \(\mathrm{SU}(2)\). The finite subgroups of \(\mathrm{SU}(2)\), their orders, generators, and notable elements are summarized in Table~\ref{table:su2-subgroup}.

\begin{table}[h!]
\centering
\begin{tabular}{c | c c c} 
     \toprule
     Group & Notable Elements & Generators & Order \\
     \midrule
     $\mathrm{C}_{2m}$ & $Z\bigl(\tfrac{2\pi}{m}\bigr)$ & $Z\bigl(\tfrac{2\pi}{m}\bigr)$ & $2m$ \\
     $\mathrm{BD}_{2m}$ & $\hat{X},\,Z\bigl(\tfrac{2\pi}{m}\bigr)$ & $\hat{X},\,Z\bigl(\tfrac{2\pi}{m}\bigr)$ & $4m$ \\
     $2\mathrm{T}$ & $\hat{X},\,\hat{Z},\,F$ & $\hat{X},\,F$ & $24$ \\
     $2\mathrm{O}$ (Clifford) & $\hat{X},\,\hat{Z},\,\hat{S},\,\hat{H},\,F$ & $\hat{S},\,\hat{H}$ & $48$ \\
     $2\mathrm{I}$ & $\hat{X},\,\hat{Z},\,F,\,\Phi$ & $\hat{X},\,\hat{Z}\Phi$ & $120$ \\
     \bottomrule
\end{tabular}
\caption{Finite subgroups of \(\mathrm{SU}(2)\), including their notable elements, minimal generating sets, and orders.}
\label{table:su2-subgroup}
\end{table}

To describe these subgroups and their properties, we rely on several standard single-qubit unitary gates, which are defined as:
\[
H = \frac{1}{\sqrt{2}}
\begin{bmatrix}
1 & 1 \\
1 & -1
\end{bmatrix}, \quad
S =
\begin{bmatrix}
1 & 0 \\
0 & i
\end{bmatrix}, \quad
Z(\theta) =
\begin{bmatrix}
e^{-i\theta/2} & 0 \\
0 & e^{i\theta/2}
\end{bmatrix}.
\]

To ensure matrices have determinant 1 (i.e., belong to \(\mathrm{SU}(2)\)), we use the "hat" notation to denote multiplication by a global factor. For example:
\[
\hat{X} = -iX, \quad \hat{Y} = -iY, \quad \hat{Z} = -iZ, \quad \hat{H} = -iH, \quad \hat{S} = Z\bigl(\tfrac{\pi}{2}\bigr) = e^{-i\pi/4}S.
\]

The unitary \(F\) and \(\Phi\) (along with its conjugate \(\Phi^\star\)) are particularly important for describing the groups \(2\mathrm{T}\), \(2\mathrm{O}\), and \(2\mathrm{I}\):
\[
F = \hat{H}Z\bigl(-\tfrac{\pi}{2}\bigr) = \frac{1}{2}
\begin{bmatrix}
1 - i & -1 - i \\
1 - i & 1 + i
\end{bmatrix}.
\]

The unitary \(\Phi\) and its conjugate \(\Phi^\star\) are defined as:
\[
\Phi = \frac{1}{2}
\begin{bmatrix}
a + i\,b & 1 \\
-1 & a - i\,b
\end{bmatrix}, \quad
\Phi^\star = \frac{1}{2}
\begin{bmatrix}
-b + i\,a & 1 \\
-1 & -b - i\,a
\end{bmatrix},
\]
where \(a = \frac{\sqrt{5} + 1}{2}\) and \(b = \frac{\sqrt{5} - 1}{2}\).

For the group \(2\mathrm{I}\), one can also use alternative rotations in the \((Y,Z)\)–plane.  Let \((p,q)\) be any integer pair satisfying
\[
  p^2 + q^2 = 5,
  \quad
  p,q \in \{\pm2,\pm1\}.
\]
Define
\[
  R_{p,q}
  \;=\;
  I\cos\!\bigl(\tfrac{\pi}{5}\bigr)
  \;+\;
  \frac{i\,\sin\!\bigl(\tfrac{\pi}{5}\bigr)}{\sqrt{5}}\,
    \bigl(p\,Y \;+\; q\,Z\bigr).
\]
Then one obtains equivalent presentations of \(2I\), for example
\[
  2I \;\cong\;
    \bigl\langle \hat X,\,\hat Z\Phi \bigr\rangle
  \;\cong\;
    \bigl\langle \hat X,\,Z(2\pi/5),\,R_{2,1}\bigr\rangle
  \;\cong\;
    \bigl\langle \hat X,\,Z(2\pi/5),\,R_{2,-1}\bigr\rangle,
\]
and so on, where each choice of \((p,q)\in\{(2,1),(2,-1),(-2,1),(-2,-1)\}\) gives an equally valid generating set.

In subsequent sections, we will see how specific quantum codes realize one or more of these finite subgroups of \(\mathrm{SU}(2)\) as their transversal logical gate set. The irreducible representations of these groups, detailed in Appendix A, are used to analyze the transversal properties of newly discovered QECCs.

\section{Methodology}
In this section, we outline the main steps for searching quantum error-correcting codes (QECCs) and their possible transversal gates. The approach builds on two core ideas: (1) parameterizing the code subspace by embedding it into a Stiefel manifold, and (2) formulating appropriate loss functions that implement the Knill-Laflamme (KL) conditions and/or the requirement of having specific transversal gates.

\subsection{Code parameterization via Stiefel manifolds}

To systematically explore potential code subspaces, we represent an \(((n,K,d))\) quantum code as a collection of $K$ orthonormal vectors in a $2^n$-dimensional Hilbert space. Concretely, let
\[ \Psi \;=\; \left(
\ket{\psi_1}, \ldots, \ket{\psi_K} \right)
\in\;\mathrm{St}(2^n,K) \;\subseteq\; \mathbb{C}^{2^n\times K},\]
where \(\mathrm{St}(m,n)\) denotes the \emph{Stiefel manifold} of all \(m\times n\) complex matrices with orthonormal columns:
\[ \mathrm{St}(m,n) =\Bigl\{ X\in \mathbb{C}^{m\times n} : X^\dagger X = I_n \Bigr\}. \]
A convenient way to map arbitrary full-rank matrices \(\Theta\in\mathbb{C}^{m\times n}\) onto this manifold is through the polar decomposition,
\[\Psi(\Theta) \;=\; \Theta\bigl(\Theta^\dagger \Theta\bigr)^{-\tfrac{1}{2}},\]
which ensures \(\Psi(\Theta)\) has orthonormal columns. Any valid code subspace of dimension $K$ in a $2^n$-dimensional Hilbert space can thus be represented by a point in \(\mathrm{St}(2^n,K)\).

\subsection{Loss function \(\mathcal{L}_{\mathrm{KL}}\) for the KL conditions}

Once the code subspace is parameterized by \(\Psi = \left(\ket{\psi_i}\right)_{i \in [K]} \in \mathrm{St}(2^n,K)\), we impose the Knill-Laflamme (KL) conditions via a \emph{loss function} that enforces correctability of Pauli errors up to a certain weight. Let \( \edset \) be the set of detectable Pauli operators (of weight less than \(d\), if searching for a code of distance \(d\)). For each error $E \in \edset$, define
\[
\tilde{\lambda}_{E, i,j}
=\langle\psi_i|\,E\,|\psi_j\rangle,\quad i,j \in [K]
\]
which are matrix entries of \(E\) in the code basis. The KL condition demands that for all \(i\neq j\), 
\(\tilde{\lambda}_{E,i,j} = 0\) and 
\(\tilde{\lambda}_{E,i,i} = \tilde{\lambda}_{E,j,j}\).  A suitable \emph{KL loss function} is
\begin{align}
\mathcal{L}_{\mathrm{KL}}\bigl(\Psi\bigr)
&:=
\sum_{E \in \edset}\sum_{i\neq j}
\;\bigl|\tilde{\lambda}_{E,i,j}\bigr|^2
\;+\;
\Bigl(\tilde{\lambda}_{E,i,i}
- \tilde{\lambda}_{E,j,j} \Bigr)^{2}.
\end{align}

Minimizing \(\mathcal{L}_{\mathrm{KL}}\) enforces the KL condition up to numerical tolerance, ensuring the resulting code subspace is a valid QECC for the specified error set \(\edset\).

\subsection{Transversal gates and their loss function \(\mathcal{L}_{\mathrm{gate}}\)}
Suppose one wants to find local unitaries \(U_1^{(r)},\ldots,U_n^{(r)}\in \mathrm{SU}(2)\) that implement certain target logical gates \(V^{(1)}, \cdots, V^{(r)} \in \mathrm{SU}(2)\), which are generators of a particular finite subgroup of $\mathrm{SU}(2)$. Then we define a \emph{gate loss function} $$\mathcal{L}_{\mathrm{gate}}: \mathrm{St}\left(2^n,2\right) \times \mathrm{SU}(2)^{n\times r} \rightarrow \mathbb{R}_{\ge 0}$$ such that
\begin{equation}
\label{eq:Lgate}
\mathcal{L}_{\mathrm{gate}}\left(\Psi,  \bigl(U^{(t)}_s\bigr)_{s\in[n], t\in[r]} \right)
=
\sum_{t=1}^r
\sum_{i,j=0}^{K-1}
\left|\,
\langle \psi_i| \bigl(\otimes_{s}U_s^{(t)}\bigr)|\psi_j\rangle
\;-\;
V^{(t)}_{ij}
\,\right|^2.
\end{equation}
Here, \(\Psi = (|\psi_i\rangle)_{i=0}^{K-1}\) represents the current code basis, and \(U_s^{(t)}\) is the local unitary on qubit \(s\) meant to realize the logical gate \(V^{(t)}\). Minimizing \(\mathcal{L}_{\mathrm{gate}}\) enforces that \(\otimes_s U_s^{(t)}\) acts on the code subspace as the logical gate \(V^{(t)}\).

\subsection{Combining loss functions for different tasks}\label{sec:combine-loss}

The framework above allows us to handle several key tasks by choosing which terms to include in the global loss function:

\begin{itemize}
\item \textbf{Finding a code only.} To find an \(((n,K,d))\) code (subspace) that satisfies the KL conditions for a target set of errors, one can simply minimize
\[ \mathcal{L}_{\mathrm{KL}}\bigl(\Psi\bigr), \]
where \(\Psi\) parameterises the subspace on \(\mathrm{St}(2^n,K)\). No gate-related constraints are imposed in this scenario.

\item \textbf{Finding a code with a prescribed signature norm.} To construct a code with a desired signature norm \(\lambda^*\), one can augment the KL loss function with a penalty term that enforces this constraint:
\[ \mathcal{L}(\Psi) = \mathcal{L}_{\mathrm{KL}}\bigl(\Psi\bigr) + \left((\tilde{\lambda}^{*})^2 - (\lambda^{*})^2\right)^{2}, \]
where \(\tilde{\lambda}^{*} = \sqrt{\sum_{E \in \mathcal{E}_{\mathrm{D}}} \left(\frac{1}{K} \sum_{i} \tilde{\lambda}_{E, i, i} \right)^{2}}\) denotes the estimated signature norm associated with the code subspace \(\Psi\).

\item \textbf{Finding a code with a prescribed transversal group.} In some cases, one aims to identify a QECC whose logical subspace admits certain transversal logical gates \(\{V^{(r)}\}\). One then combines the KL condition and the gate conditions into a single loss:
\[
\mathcal{L}(\Psi)
=
\mathcal{L}_{\mathrm{KL}}(\Psi)
\;+\;
\mathcal{L}_{\mathrm{gate}}\left(\Psi, \bigl(U^{(t)}_s\bigr)_{s\in[n], t\in[r]} \right),
\]
thus simultaneously enforcing correctability of the desired errors and ensuring the subspace supports the specified transversal group actions.

\item \textbf{Finding transversal implementations for a \emph{known} code.} If a code (subspace) is already fixed, we keep \(\{|\psi_i\rangle\}\) fixed and optimize only the unitaries $\{\otimes_s U^{(r)}_s\}$. In this case, we use
\[
\mathcal{L}_{\mathrm{gate}}\left(  \bigl(U^{(t)}_s\bigr)_{s\in[n], t\in[r]}\right)
\;=\;
\sum_{r=1}^R
\sum_{i,j=0}^{K-1}
\bigl|\,
\langle \psi_i|\otimes_s U_s^{(r)}|\psi_j\rangle
-\,
V^{(r)}_{ij}\bigr|^2.
\]
Minimizing \(\mathcal{L}_{\mathrm{gate}}\) leads to local unitaries implementing the given logical gates \(\{V^{(r)}\}\) transversally on that code.
\end{itemize}

In all cases, the search can be conducted using gradient-based methods on the relevant manifolds. For instance, when optimizing single-qubit unitaries \(U_s\in \mathrm{SU}(2)\), one can map an unconstrained vector \(\theta\in \mathbb{R}^3\) to \(\mathrm{SU}(2)\) via matrix exponentials or other trivialization techniques~\cite{lezcano2019trivializations}, thereby turning the constrained optimization into an unconstrained problem on \(\mathbb{R}^3\). Similarly, the subspace \(\Psi \in\mathrm{St}(2^n,K)\) can be parameterized via the polar decomposition as discussed above. Numerical minimization then proceeds (for example) by alternating or jointly optimizing \(\Psi\) (which defines the code subspace) and the local unitaries \(\{U_s\}\).

By choosing suitable loss functions and parameterizations, one can thus efficiently search for QECCs, evaluate the code's transversal gate group, or find subspaces that realize specific target transversal operators.

\section{Transversal Groups for \texorpdfstring{$((6,2,3))$}{((6,2,3))} Codes}

Unlike the \(((5,2,3))\) code, which is unique up to local unitary equivalence~\cite{rains1999quantumCodes}, the \(((6,2,3))\) code space admits a richer variety of transversal gate groups. We systematically explore this space by applying our composite loss function across candidate groups, aiming to identify logical subspaces that simultaneously satisfy the Knill–Laflamme conditions and support the desired transversal symmetry. The results are summarized in Table~\ref{table:623-trans-group}.

For several groups—including $\mathrm{C}_2$, $\mathrm{C}_4$, $\mathrm{C}_6$, $\mathrm{C}_{10}$, $\mathrm{BD}_4$, and $2\mathrm{T}$—our optimization procedure achieves near-zero loss, indicating the existence of codes that   these transversal groups. In contrast, for other candidates such as $\mathrm{C}_{12}$, $\mathrm{C}_{14}$, and $\mathrm{BD}_8$, the loss remains significantly nonzero, suggesting either the absence of compatible codes or the presence of suboptimal local minima in our search.

To further characterize the identified codes, we compute their quantum weight enumerators. Remarkably, the results cluster into two parametric families and one isolated point:
\begin{align}
    A &= \left[1,a,1-a,0,11+4a,16-a,3-3a\right], && a \in [0,1], \label{eq:623-shorwt-SU4} \\
    A &= \left[1,0,\tfrac{1+a}{2},\tfrac{1}{2}+a,\tfrac{23}{2}-2a,\tfrac{31}{2}-a,3\right], && a \in [0.2,1], \label{eq:623-shorwt-SO5} \\
    A &= \left[1,0,0.84,0,11.64,15.36,3.16\right]. \label{eq:623-shorwt-C10}
\end{align}
These weight enumerators provide a useful way to distinguish code classes and are directly related to the structure of their corresponding transversal groups. Although we cannot completely rule out additional classes due to possible local minima in our optimization process, our numerical results strongly indicate that transversal groups combined with quantum weight enumerators effectively classify \(((6,2,3))\) codes.

\begin{table}[h!]
\centering
\begin{tabular}{c | c || c | c}
    \toprule
    Group & $\lambda^*$ Range & Group & $\lambda^*$ Range \\
    \midrule
    $2\mathrm{O}$ & N/A & $\mathrm{C}_{12}$ & N/A \\
    $2\mathrm{I}$ & N/A & $\mathrm{C}_{14}$ & N/A \\
    $2\mathrm{T}$ & $\{1\}$ & $\mathrm{C}_{16}$ & N/A \\
    $\mathrm{C}_2$ & $[\sqrt{0.6},1]$ & $\mathrm{C}_{18}$ & N/A \\
    $\mathrm{C}_4$ & $[\sqrt{2/3},1]$ & $\mathrm{BD}_4$ & $[\sqrt{3/4},1]$ \\
    $\mathrm{C}_6$ & $\{1\}$ & $\mathrm{BD}_6$ & N/A \\
    $\mathrm{C}_8$ & N/A & $\mathrm{BD}_8$ & N/A \\
    $\mathrm{C}_{10}$ & $\{\sqrt{0.84}\}$ & & \\
    \bottomrule
\end{tabular}
\caption{Transversal groups for the \(((6,2,3))\) quantum code. The "\(\lambda^*\) Range" column reports the numerically optimized values of the signature norm \(\lambda^*\) associated with each group, indicating whether a valid code with the specified transversal group was found. A value of "N/A" indicates that no such code has been found so far, while exact or interval values of \(\lambda^*\) correspond to successful constructions.}
\label{table:623-trans-group}
\end{table}

In the following, we highlight a novel \(((6,2,3))\) quantum code whose transversal group is isomorphic to \(\mathrm{C}_{10}\). Unlike earlier constructions based on extensions of the \(((5,2,3))\) code~\cite{Cao_2022} or \(\mathrm{SO}(5)\) parametrizations~\cite{du2024characterizing} (which are summarized in Appendix~\ref{appendix:623families}), this new code achieves a genuinely non-Clifford transversal gate structure. In particular, while transversal groups such as \(\mathrm{C}_2\), \(\mathrm{C}_4\), and \(\mathrm{BD}_4\) can arise from Clifford-symmetric constructions, and \(\mathrm{C}_6\) is a subgroup of \(2\mathrm{T}\) accessible via local unitary rotations, the realization of \(\mathrm{C}_{10}\) marks a fundamental departure. To our knowledge, this \(((6,2,3))\) code represents the smallest known example supporting a non-Clifford transversal gate, highlighting a significant advancement in the exploration of low-dimensional nonadditive quantum codes.

\subsection{\texorpdfstring{$((6,2,3))$}{((6,2,3))} Code with Transversal Group \texorpdfstring{$\mathrm{C}_{10}$}{C₁₀}}

We conclude this section by presenting a \(((6,2,3))\) quantum code that admits a non-Clifford transversal gate, with its transversal group isomorphic to \(\mathrm{C}_{10}\). This construction represents the smallest known quantum code supporting a transversal implementation of a gate outside the Clifford group. The logical basis states are given by:
\begin{equation}\label{eq:623C10}
\begin{aligned}
    \sqrt{30}\left|0_{L}\right\rangle = & \sqrt{3}\left|000000\right\rangle +\sqrt{3}e^{i\theta_{1}}\left|000011\right\rangle +\sqrt{3}e^{i\theta_{2}}\left|011110\right\rangle +\sqrt{3}e^{i\theta_{3}}\left|101110\right\rangle \\
    & +\sqrt{3}e^{i\theta_{4}}\left|110110\right\rangle +\sqrt{3}e^{i\theta_{5}}\left|111010\right\rangle +\sqrt{2}e^{i\theta_{6}}\left|001101\right\rangle +\sqrt{2}e^{i\theta_{7}}\left|010101\right\rangle \\
    & +\sqrt{2}e^{i\theta_{8}}\left|011001\right\rangle +\sqrt{2}e^{i\theta_{9}}\left|100101\right\rangle +\sqrt{2}e^{i\theta_{10}}\left|101001\right\rangle +\sqrt{2}e^{i\theta_{11}}\left|110001\right\rangle \\
    \sqrt{30}\left|1_{L}\right\rangle = & \sqrt{3}\left|111111\right\rangle +\sqrt{3}e^{-i\theta_{1}}\left|111100\right\rangle -\sqrt{3}e^{-i\theta_{2}}\left|100001\right\rangle -\sqrt{3}e^{-i\theta_{3}}\left|010001\right\rangle \\
    & -\sqrt{3}e^{-i\theta_{4}}\left|001001\right\rangle -\sqrt{3}e^{-i\theta_{5}}\left|000101\right\rangle -\sqrt{2}e^{-i\theta_{6}}\left|110010\right\rangle -\sqrt{2}e^{-i\theta_{7}}\left|101010\right\rangle \\
    & -\sqrt{2}e^{-i\theta_{8}}\left|100110\right\rangle -\sqrt{2}e^{-i\theta_{9}}\left|011010\right\rangle -\sqrt{2}e^{-i\theta_{10}}\left|010110\right\rangle -\sqrt{2}e^{-i\theta_{11}}\left|001110\right\rangle 
\end{aligned}
\end{equation}
The phase parameters \(\{\theta_j\}_{j=1}^{11}\) satisfy the following constraints:
\[ \begin{cases}
e^{i\theta_{5}+i\theta_{9}}+e^{i\theta_{4}+i\theta_{10}}+e^{i\theta_{3}+i\theta_{11}}=0,\\
e^{i\theta_{5}+i\theta_{7}}+e^{i\theta_{4}+i\theta_{8}}+e^{i\theta_{2}+i\theta_{11}}=0,\\
e^{i\theta_{5}+i\theta_{6}}+e^{i\theta_{3}+i\theta_{8}}+e^{i\theta_{2}+i\theta_{10}}=0,\\
e^{i\theta_{4}+i\theta_{6}}+e^{i\theta_{3}+i\theta_{7}}+e^{i\theta_{2}+i\theta_{9}}=0,\\
e^{i\theta_{6}+i\theta_{11}}+e^{i\theta_{7}+i\theta_{10}}+e^{i\theta_{8}+i\theta_{9}}=0.
\end{cases} \]
These equations are invariant under local Pauli \(Z\) rotations. For example, applying a \(Z(\alpha)\) rotation to the first qubit uniformly introduces a phase \(e^{i\alpha}\) to each term in the first equation, which cancels when summed. Thus, such local phases can be fixed to simplify the solution space. Using this freedom, we set six phase parameters to zero, leaving only two distinct solutions:
\[ \vec{\theta} = \left(0, 0, \tfrac{2\pi}{3}, \tfrac{4\pi}{3}, 0, 0, 0, 0, 0, \tfrac{4\pi}{3}, \tfrac{2\pi}{3}\right), \quad
\vec{\theta} = \left(0, 0, \tfrac{4\pi}{3}, \tfrac{2\pi}{3}, 0, 0, 0, 0, 0, \tfrac{2\pi}{3}, \tfrac{4\pi}{3}\right).\]
These two codewords are related by the local unitary operator \(X \otimes X \otimes X \otimes X \otimes Y \otimes Y\), and are therefore equivalent under local unitary transformations. Hence, all codes described by Equation~\eqref{eq:623C10} form a single LU-equivalence class.

The code admits a transversal non-Clifford logical operator given by:
\[ \overline{Z\left(-\tfrac{2\pi}{5}\right)}=Z\left(\tfrac{2\pi}{5}\right)\otimes Z\left(\tfrac{2\pi}{5}\right)\otimes Z\left(\tfrac{2\pi}{5}\right)\otimes Z\left(\tfrac{2\pi}{5}\right)\otimes Z\left(\tfrac{4\pi}{5}\right)\otimes Z\left(\tfrac{6\pi}{5}\right), \]
which generates a cyclic group of order 10. This makes the code the first \(((6,2,3))\) construction known to realize \(\mathrm{C}_{10}\) as its transversal gate group. The quantum weight enumerator of this code is given by Equation~\eqref{eq:623-shorwt-C10}, and is distinct from those of all previously discussed \(((6,2,3))\) codes, further supporting its inequivalence under local unitaries.

\section{\texorpdfstring{$((7,2,3))$}{((7,2,3))} Codes with Exceptional Transversal Groups}

Beyond the \(((6,2,3))\) codes, the \(((7,2,3))\) code family offers a significantly larger design space and potential for realizing richer transversal gate symmetries. We examine all \(((7,2,3))\) stabilizer codes, which have been fully classified in~\cite{cross2025small}, and evaluate their transversal groups using our optimization-based search algorithm. Our analysis reveals the following possibilities: the binary octahedral group (\(2\mathrm{O}\), e.g., the Steane code~\cite{steane1996simple}), the binary tetrahedral group (\(2\mathrm{T}\), obtainable by appending two ancilla qubits in the \(\ket{0}\) state to the \(((5,2,3))\) code), \(\mathrm{BD}_8\) (specifically, code index 209 in~\cite{cross2025small}), and \(\mathrm{BD}_4\) (found in all other stabilizer codes).

We then extend our search beyond stabilizer codes by applying the composite loss function introduced in Section~\ref{sec:combine-loss}. This function combines the Knill–Laflamme  loss, a signature norm penalty, and a gate loss that favors specific transversal group actions. For each group of interest, we optimize over code subspaces to identify candidates with the desired symmetry and explore their associated signature norms \(\lambda^*\). Our findings are summarized in Table~\ref{table:723-transversal-groups}. In addition to the groups found in stabilizer codes, our search uncovers \(((7,2,3))\) codes that support transversal groups isomorphic to the binary icosahedral group (\(2\mathrm{I}\)) \cite{kubischta2023family} and a broad family of binary dihedral groups \(\mathrm{BD}_{2m}\) with \(2m \in [12,36]\). For these codes, the signature norm \(\lambda^*\) varies within a wide range and serves as a useful invariant for classification. We also observe that the quantum weight enumerators for many \(((7,2,3))\) codes align along specific parametric lines, given by:
\begin{equation}\label{eq:723shorwt-line}
\begin{aligned}
A &= \left[1, 0, \lambda^{*2}, 0, 21 - 2\lambda^{*2}, 0, 42 + \lambda^{*2}, 0\right], \\
B &= \left[1, 0, \lambda^{*2}, 21 + 3\lambda^{*2}, 21 - 2\lambda^{*2}, 126 - 6\lambda^{*2}, 42 + \lambda^{*2}, 45 + 3\lambda^{*2}\right],
\end{aligned}
\end{equation}
where \(\lambda^* \in [0, \sqrt{7}]\). These expressions capture a broad class of codes, including several known stabilizer examples from~\cite{cross2025small}, as well as newly discovered nonadditive constructions.

\begin{table}[h!]
\centering
\begin{tabular}{c | c || c | c}
    \toprule
    \textbf{Group} & \(\lambda^*\) Range & \textbf{Group} & \(\lambda^*\) Range \\
    \midrule
    \(2\mathrm{O}\) & \([0, 2]\) & \(\mathrm{BD}_{24}\) & \([1, \sqrt{3}]\) \\
    \(2\mathrm{I}\) & \(\{0, \sqrt{3/4}, \sqrt{7}\} \cup [1.34, 1.52]\) & \(\mathrm{BD}_{26}\) & \([1.13, 1.83]\) \\
    \(2\mathrm{T}\) & \([0, \sqrt{7}]\) & \(\mathrm{BD}_{28}\) & \([1.13, 1.68]\) \\
    \(\mathrm{BD}_6\) & \([0, \sqrt{7}]\) & \(\mathrm{BD}_{30}\) & \([1.47, 1.71]\) \\
    \(\mathrm{BD}_8\) & \([0, 2.27]\) & \(\mathrm{BD}_{32}\) & \([1.29, 1.44]\) \\
    \(\mathrm{BD}_{10}\) & \([0, 2.52] \cup \{\sqrt{7}\}\) & \(\mathrm{BD}_{34}\) & \(\Bigl\{ \sqrt{\tfrac{681}{289}}, \sqrt{\tfrac{831}{289}} \Bigr\}\) \\
    \(\mathrm{BD}_{12}\) & \([0, 2.25]\) & \(\mathrm{BD}_{36}\) & \(\Bigl\{ \sqrt{\tfrac{161}{81}} \Bigr\}\) \\
    \(\mathrm{BD}_{14}\) & \([0.67, 2.17]\) & \(\mathrm{BD}_{38}\) & N/A \\
    \(\mathrm{BD}_{16}\) & \([0.77, 2.2]\) & \(\mathrm{BD}_{40}\) & N/A \\
    \(\mathrm{BD}_{18}\) & \([0.77, 2.12]\) & \(\mathrm{C}_{38}\) & N/A \\
    \(\mathrm{BD}_{20}\) & \([0.82, 1.87]\) & \(\mathrm{C}_{40}\) & N/A \\
    \(\mathrm{BD}_{22}\) & \([1.06, 1.85]\) &  &  \\
    \bottomrule
\end{tabular}
\caption{Transversal groups discovered for \(((7,2,3))\) quantum codes. The \(\lambda^*\) range indicates the numerically estimated interval of signature norms that arise for codes admitting the specified group. "N/A" denotes groups for which no code was found under our current search. Note that the listed intervals are not exhaustive but represent all ranges discovered so far.}
\label{table:723-transversal-groups}
\end{table}

In this section, we highlight new \(((7,2,3))\) quantum error-correcting codes that admit a transversal group isomorphic to the binary icosahedral group \(2\mathrm{I}\), an exceptional finite subgroup of \(\mathrm{SU}(2)\). Constructions realizing \(2\mathrm{T}\) and \(2\mathrm{O}\) transversal symmetry are summarized in Appendix~\ref{appendix:723exceptional}. In contrast, the new \(((7,2,3))\) codes presented here achieve \(2\mathrm{I}\) transversal symmetry with novel parameter regimes, including cases with signature norms \(\lambda^* = 0\) and \(\lambda^* = \sqrt{3/4}\). These constructions complement the previously known permutation-invariant code~\cite{pollatsek2004permutationally,kubischta2023family} with \(\lambda^* = \sqrt{7}\). Transversal groups of the form \(\mathrm{BD}_{2m}\) and \(\mathrm{C}_{2m}\) will be discussed in detail in the subsequent section.

\subsection{\texorpdfstring{$((7,2,3))$}{((7,2,3))} Code with Transversal Group \texorpdfstring{$2\mathrm{I}$}{2I}}

A \(((7,2,3))\) code with transversal symmetry group isomorphic to the binary icosahedral group \(2\mathrm{I}\) was introduced in~\cite{kubischta2023family}. This code is permutation-invariant and its logical basis states are defined using symmetric Dicke states:
\[ |0_L\rangle = \frac{\sqrt{3}\,|D^7_0\rangle + i\sqrt{7}\,|D^7_5\rangle}{\sqrt{10}}, \quad
|1_L\rangle = \frac{i\sqrt{7}\,|D^7_2\rangle + \sqrt{3}\,|D^7_7\rangle}{\sqrt{10}}. \]
where \(\ket{D^n_k}\) denotes the symmetric Dicke state of \(n\) qubits with Hamming weight \(k\) \cite{PhysRev.93.99},
\[ |D^n_k\rangle \;=\;\binom{n}{k}^{-\frac12} \sum_{\substack{x\in\{0,1\}^n\\|x|=k}}|x\rangle. \]
This code admits the following transversal logical gates:
\[ \overline{X} = X^{\otimes7},  \quad
\overline{Z\left(\tfrac{2\pi}{5}\right)} = \bigl(Z(\tfrac{6\pi}{5})\bigr)^{\otimes7},  \quad
\overline{R_{-2,1}} = \bigl(-R_{-2,-1}^{3}\bigr)^{\otimes7}, \]
which generate the group \(2\mathrm{I}\). Notably, this code is local unitary equivalent to the cyclic \(((7,2,3))\) code with \(\lambda^* = \sqrt{7}\) defined in Equation~\eqref{eq:723cyclic}.

In addition to this previously known example, we present two new \(((7,2,3))\) codes with transversal group \(2\mathrm{I}\), exhibiting signature norms \(\lambda^* = 0\) and \(\lambda^* = \sqrt{3/4}\), respectively.

\paragraph{Code with \texorpdfstring{$\lambda^* = 0$}{λ* = 0}.}This code is constructed within a 4-dimensional subspace spanned by orthonormal vectors \(\{\ket{S_i}\}_{i=0}^3\), defined as:
\begin{align*}
    \sqrt{8}\left|S_{0}\right\rangle =&\left|0000000\right\rangle +\left|0111111\right\rangle -\left|1011111\right\rangle +\left|1101111\right\rangle -\left|1110001\right\rangle +\left|1110010\right\rangle +\left|1110100\right\rangle +\left|1111000\right\rangle ,\\
    \sqrt{6}\left|S_{1}\right\rangle =&-\left|0010011\right\rangle +\left|0011100\right\rangle +\left|0100101\right\rangle -\left|0101010\right\rangle +\left|1000110\right\rangle -\left|1001001\right\rangle ,\\
    \sqrt{6}\left|S_{2}\right\rangle =&\left|0010101\right\rangle -\left|0011010\right\rangle +\left|0100110\right\rangle -\left|0101001\right\rangle +\left|1000011\right\rangle -\left|1001100\right\rangle ,\\
    \sqrt{6}\left|S_{3}\right\rangle =&-\left|0010110\right\rangle +\left|0011001\right\rangle -\left|0100011\right\rangle +\left|0101100\right\rangle +\left|1000101\right\rangle -\left|1001010\right\rangle .
\end{align*}
The logical basis states are then given by:
\begin{align*}
    \left|0_{L}\right\rangle =&c_{0}\left|S_{0}\right\rangle +c_{1}\left|S_{1}\right\rangle +c_{2}\left|S_{2}\right\rangle +c_{3}\left|S_{3}\right\rangle ,\\
    \left|1_{L}\right\rangle =&X^{\otimes7}\left|0_{L}\right\rangle,
\end{align*}
with coefficients \(c_0, c_1, c_2, c_3\) depending on real parameters \(\theta, \phi\) and sign choice:
\[ \begin{cases}
c_{0}=\pm\sqrt{\frac{2}{5}},\\
c_{1}=-\sqrt{\frac{1}{3}}\sin(\theta)\cos(\phi)+i\sqrt{\frac{1}{3}}\cos(\theta)\left(\cos(\phi-\pi/3)+\cos(\phi+\pi/3)\right)+i\sqrt{\frac{1}{30}},\\
c_{2}=-\sqrt{\frac{1}{3}}\sin(\theta)\cos(\phi-\pi/3)+i\sqrt{\frac{1}{3}}\cos(\theta)\cos(\phi-\pi/3)-i\sqrt{\frac{1}{30}},\\
c_{3}=-\sqrt{\frac{1}{3}}\sin(\theta)\left(\cos(\phi-\pi/3)-\cos(\phi)\right)+i\sqrt{\frac{1}{3}}\cos(\theta)\cos(\phi+\pi/3)-i\sqrt{\frac{1}{30}}.
\end{cases} \]
Three generators of the transversal group \(2\mathrm{I}\) can be realized as:
\[ \overline{X}=X^{\otimes7}, \]
\[ \overline{Z\left(\tfrac{2\pi}{5}\right)}=Z\left(\tfrac{2\pi}{5}\right)\otimes Z\left(\tfrac{2\pi}{5}\right)\otimes Z\left(\tfrac{2\pi}{5}\right)\otimes Z\left(\tfrac{4\pi}{5}\right)\otimes Z\left(\tfrac{4\pi}{5}\right)\otimes Z\left(\tfrac{4\pi}{5}\right)\otimes Z\left(\tfrac{4\pi}{5}\right). \]
The third generator depends on the sign of \(c_0\). When \(c_0 = \sqrt{\tfrac{2}{5}}\), we have:
\[ \overline{R_{-2,1}}=R_{2,1}\otimes R_{-2,1}\otimes R_{2,1}\otimes R_{2,-1}^{2}\otimes R_{2,-1}^{2}\otimes R_{2,-1}^{2}\otimes R_{-2,-1}^{2}, \]
and when $c_{0}=-\sqrt{\frac{2}{5}}$,
\[ \overline{R_{2,1}}=R_{-2,1}\otimes R_{2,1}\otimes R_{-2,1}\otimes R_{-2,-1}^{2}\otimes R_{-2,-1}^{2}\otimes R_{-2,-1}^{2}\otimes R_{2,-1}^{2}. \]

It is worth noting that the first two generators are automatically satisfied within this subspace; the enforcement of the Knill–Laflamme condition depends on the third generator and the chosen coefficients. For instance, the subspace spanned by \(\{\ket{S_1} + \ket{S_2}, X^{\otimes 7}(\ket{S_1} + \ket{S_2})\}\) also supports a \(2\mathrm{I}\) transversal structure but fails to meet the QEC distance \(d=3\) constraint.

The quantum weight enumerators for this family of codes are parametrized by \(\theta\) as:
\[ A=\left[1,0,0,0,21-8\sin^{2}\theta,24\sin^{2}\theta,42-24\sin^{2}\theta,8\sin^{2}\theta\right], \]
\[ B=\left[1,0,0,21-8,21+32\sin^{2}\theta,126-48\sin^{2}\theta,42+32\sin^{2}\theta,45-8\sin^{2}\theta\right].\]
These expressions imply that codes with different values of \(\sin^2\theta\) are not locally unitary equivalent. Additionally, numerical evidence suggests that even codes with the same weight enumerator but different \(\phi\) may be LU inequivalent.

\paragraph{Code with \texorpdfstring{$\lambda^* = \sqrt{3/4}$}{λ* = √(3/4)}.}We now present a second \(((7,2,3))\) code with transversal group \(2\mathrm{I}\), for which the signature norm takes the value \(\lambda^* = \sqrt{3/4}\). The logical subspace is embedded in a four-dimensional space spanned by the orthonormal vectors \(\{\ket{S_0}, \ket{S_1}, \ket{S_2}, \ket{S_3}\}\), defined as:
\begin{align*}
    \left|S_{0}\right\rangle =&\left|0000000\right\rangle ,\\
    \sqrt{6}\left|S_{1}\right\rangle =&\left|0000111\right\rangle -\left|0001011\right\rangle +\left|0010011\right\rangle +\left|0100011\right\rangle +\left|1000011\right\rangle -\left|1111100\right\rangle, \\
    \sqrt{10}\left|S_{2}\right\rangle =&\left|0011101\right\rangle +\left|0101101\right\rangle -\left|0110110\right\rangle +\left|0111010\right\rangle +\left|1001110\right\rangle \\&-\left|1010101\right\rangle +\left|1011010\right\rangle -\left|1100110\right\rangle +\left|1101001\right\rangle -\left|1110001\right\rangle, \\
    \sqrt{10}\left|S_{3}\right\rangle =&\left|0011110\right\rangle +\left|0101110\right\rangle -\left|0110101\right\rangle +\left|0111001\right\rangle +\left|1001101\right\rangle \\&-\left|1010110\right\rangle +\left|1011001\right\rangle -\left|1100101\right\rangle +\left|1101010\right\rangle -\left|1110010\right\rangle.
\end{align*}
Logical codewords are expressed as:
\begin{align*}
    \left|0_{L}\right\rangle =&c_{0}\left|S_{0}\right\rangle +c_{1}\left|S_{1}\right\rangle +c_{2}\left|S_{2}\right\rangle +c_{3}\left|S_{3}\right\rangle, \\
    \left|1_{L}\right\rangle =&X^{\otimes7}\left|0_{L}\right\rangle, 
\end{align*}
where the coefficients are given by:
\[ \begin{cases}
c_{0}=\frac{1}{\sqrt{10}},\\
c_{1}=is_{1}\sqrt{\frac{3}{20}},\\
c_{2}=\frac{s_{2}}{4}-t-is_{3}\sqrt{\frac{5}{16}-t^{2}},\\
c_{3}=\frac{s_{2}}{4}+t+is_{3}\sqrt{\frac{5}{16}-t^{2}},
\end{cases} \]
with \(s_1, s_2, s_3 \in \{\pm 1\}\) and \(t^2 \in [0, \tfrac{5}{16}]\). Regardless of the specific values chosen for the parameters, the code admits the transversal gate:
\[ \overline{Z\left(-\tfrac{2\pi}{5}\right)}=Z\left(\tfrac{2\pi}{5}\right)\otimes Z\left(\tfrac{2\pi}{5}\right)\otimes Z\left(\tfrac{2\pi}{5}\right)\otimes Z\left(\tfrac{2\pi}{5}\right)\otimes Z\left(\tfrac{2\pi}{5}\right)\otimes Z\left(\tfrac{4\pi}{5}\right)\otimes Z\left(\tfrac{4\pi}{5}\right).\]
To complete the group \(2\mathrm{I}\), a second non-Clifford transversal generator is implemented depending on the signs of \(s_1\) and \(s_2\):\\
When \(s_1 = s_2 = 1\),
\[ \overline{R_{-2,1}}=R_{2,-1}\otimes R_{2,-1}\otimes R_{2,-1}\otimes R_{-2,-1}\otimes R_{2,-1}\otimes R_{-2,1}^{2}\otimes R_{-2,1}^{2}, \]
when \(s_1 = 1\), \(s_2 = -1\),
\[ \overline{R_{-2,1}}=R_{2,-1}\otimes R_{2,-1}\otimes R_{2,-1}\otimes R_{-2,-1}\otimes R_{2,-1}\otimes R_{2,1}^{2}\otimes R_{2,1}^{2}, \]
when \(s_1 = -1\), \(s_2 = 1\),
\[ \overline{R_{2,1}}=R_{-2,-1}\otimes R_{-2,-1}\otimes R_{-2,-1}\otimes R_{2,-1}\otimes R_{-2,-1}\otimes R_{2,1}^{2}\otimes R_{2,1}^{2}, \]
when \(s_1 = s_2 = -1\),
\[ \overline{R_{2,1}}=R_{-2,-1}\otimes R_{-2,-1}\otimes R_{-2,-1}\otimes R_{2,-1}\otimes R_{-2,-1}\otimes R_{-2,1}^{2}\otimes R_{-2,1}^{2}. \]

To confirm the signature norm, we observe that the two-qubit reduced density matrices satisfy:
\[ \left\langle I^{\otimes 5}\otimes X\otimes X\right\rangle =\left\langle I^{\otimes 5}\otimes Y\otimes Y\right\rangle =\left\langle I^{\otimes 5}\otimes Z\otimes Z\right\rangle =-\frac{1}{2}, \]
implying that \(\lambda^* = \sqrt{3/4}\). The quantum weight enumerators for this code family depend linearly on the parameter \(t^2\), and are given by:
\[ A=\left[1,0,\frac{3}{4},0,12+24t^{2},\frac{45}{2}-72t^{2},\frac{81}{4}+72t^{2},\frac{15}{2}-24t^{2}\right], \]
\[ B=\left[1,0,\frac{3}{4},\frac{63}{4}+24t^{2},\frac{99}{2}096t^{2},\frac{153}{2}+144t^{2},\frac{291}{4}-96t^{2},\frac{159}{4}+24t^{2}\right]. \]

\section{Searching for Codes with Transversal Diagonal Gates via the SS-LP Method}

We introduce the \emph{Subset-Sum–Linear-Programming} (SS-LP) method, a systematic approach for constructing non-additive $((n,K,3))$ quantum codes that admit a transversal logical diagonal gate of the form
\[
D = \sum_{k=0}^{K-1} e^{2\pi i b_k/m}\ket{k}\bra{k},
\]
where $0 \le b_k < m$ are distinct integers. Without loss of generality, we set $b_0 = 0$ and assume the logical state $\ket{0_L}$ has nonzero overlap with the computational basis state $\ket{0^n}$.
We further assume that the logical gate $D$ is realized transversally as
\[
\overline{D} = \bigotimes_{j=1}^n \left(\ket{0}\bra{0} + e^{2\pi i a_j/m}\ket{1}\bra{1}\right),
\]
where the integers $0 \le a_1 \le \dots \le a_n < m$ define the \emph{angle vector} $\vec{a}$. For this transversal implementation to be consistent, the logical code states must satisfy
\[
\bra{k_L}\overline{D}\ket{k'_L} = e^{2\pi i b_k/m}\delta_{kk'}, \quad\text{for all } k,k'.
\]

In particular, for the special case where $K=2$ and $b_1 = m-1$, we have
\[
D = e^{-i\pi/m} Z(-2\pi/m),
\]
and the resulting code admits a transversal symmetry corresponding to the cyclic group $C_{2m}$.

\subsection{SS-LP Method}
Our SS-LP method leverages these consistency conditions to systematically search for viable codes. It consists of three main steps:
\begin{enumerate}
    \item For each logical basis state, construct its supporting subset of computational basis states by imposing the congruence condition defined by the transversal gate angles.
    \item Efficiently filter these subsets using a linear-programming feasibility test derived from $Z$-type KL conditions.
     \item Perform block-separable non-convex optimization independently within each subset \( S_k \). Since each logical basis state \(\ket{k_L}\) is supported exclusively on its subset \( S_k \), this significantly reduces the search space compared to optimizing over the entire Hilbert space.
    
\end{enumerate}

\paragraph{Step 1: Determining Logical-State Support Subsets.}
Expressing each logical basis state as $\ket{k_L} = \sum_{s\in\{0,1\}^n} c_s^{(k)}\ket{s}$, consistency under the transversal action implies that every computational basis state $\ket{s}$ appearing with nonzero amplitude $c_s^{(k)}$ must satisfy the congruence
\[
\sum_{j=1}^n a_j s_j \equiv b_k \pmod{m}.
\]
Thus, each logical state $\ket{k_L}$ is supported precisely on the subset
\[
S_k := \left\{ s\in\{0,1\}^n : \sum_{j=1}^n a_j s_j \equiv b_k \pmod{m}\right\}.
\]

Since the integers $b_k$ are distinct modulo $m$, these subsets $S_k$ are disjoint. This ensures orthogonality of distinct logical states: $\braket{k_L|k_L'} = 0$ for $k\ne k'$. Typically, the subsets $S_k$ are much smaller than the full computational basis, significantly simplifying the search. Nonetheless, the number of possible angle vectors $\vec{a}$ remains large, motivating efficient filtering methods.

\paragraph{Step 2: Linear-Programming Filter from Z-type KL Conditions.}
In principle, a distance-3 quantum code must satisfy Knill–Laflamme (KL) conditions arising from all single- and two-qubit Pauli errors. However, KL conditions associated with $X$ and $Y$ errors involve matrix elements connecting distinct subsets $S_k$, creating nonlinear constraints that are computationally demanding to handle directly. Therefore, in this initial filtering stage, we focus exclusively on the simpler diagonal KL conditions resulting from single- and two-qubit Pauli-$Z$ errors. We defer verification of the remaining KL conditions to a subsequent nonlinear optimization stage.

Explicitly, the diagonal KL conditions require consistency of logical expectation values of operators $Z_i$ and $Z_iZ_j$ across all logical states:
\[
\langle k_L|Z_i|k_L\rangle = \alpha_i,\quad \langle k_L|Z_iZ_j|k_L\rangle = \beta_{ij},\quad\forall\,k\in\{0,\dots,K-1\},\, 1\le i<j\le n,
\]
for real constants $\alpha_i$ and $\beta_{ij}$ independent of the logical index $k$. (Note that off-diagonal KL conditions vanish automatically due to disjoint supports.)

Introducing probability variables $x_s^{(k)} = |c_s^{(k)}|^2$, we formulate these constraints as a linear feasibility problem involving auxiliary variables $\alpha_i$ and $\beta_{ij}$:
\[
\begin{aligned}
&\sum_{s\in S_k}(-1)^{s_i} x_s^{(k)} = \alpha_i,\quad 0\le k < K,\; 1\le i \le n;\\[4pt]
&\sum_{s\in S_k}(-1)^{s_i+s_j} x_s^{(k)} = \beta_{ij},\quad 0\le k < K,\;1\le i < j\le n;\\[4pt]
&\sum_{s\in S_k} x_s^{(k)} = 1,\quad 0\le k < K;\\[4pt]
&x_s^{(k)}\ge 0,\quad 0\le k < K,\; s\in S_k.
\end{aligned}
\]

If this linear program (LP) is infeasible for a given angle vector $\vec{a}$ (and corresponding subsets $S_k$), we discard this choice immediately, avoiding the more computationally demanding nonlinear optimization over the full set of amplitudes $c_s^{(k)}$.

\paragraph{Step 3: Block-Separable Optimization within Reduced Subspaces.}
For each subset \( S_k \) identified and retained from Steps 1 and 2, we optimize the amplitudes \( c_s^{(k)} \) of the logical state \(\ket{k_L}\) (with \( s \in S_k \)) to minimize the full KL-loss function defined in Section~\ref{sec:combine-loss}. Since each logical state is strictly confined to its subset \( S_k \), the optimization naturally separates into smaller, independent problems, greatly simplifying the numerical search compared to a full Hilbert space optimization.

\subsection{Searching for \texorpdfstring{$((n,2,3))$}{((n,2,3))} Codes with Transversal \texorpdfstring{$\mathrm{BD}_{2m}$}{BD₂ₘ} Symmetry}

We now specialize the SS-LP method to systematically search for $((n,2,3))$ codes admitting a transversal symmetry described by the binary dihedral group $\mathrm{BD}_{2m}$. This scenario is particularly interesting for two reasons: (i) Finding a code with $\mathrm{BD}_{2m}$ symmetry automatically implies the existence of a simpler cyclic symmetry $C_{2m}$; and (ii) The richer symmetry structure provided by $\mathrm{BD}_{2m}$ further constrains the search space, significantly simplifying numerical computations.

In this setting, we assume the logical bit-flip operator is implemented transversally as $\overline{X} = X^{\otimes n}$, directly relating the logical basis states through bitwise complementation, $\ket{1_L} = X^{\otimes n}\ket{0_L}$. Additionally, requiring a nonzero overlap with the computational basis states $\ket{0^n}$ and $\ket{1^n}$ yields a congruence constraint on the angle vector $\vec{a}$:
\begin{equation}\label{eq:a-congruence}
    \sum_{j=1}^n a_j \equiv -1 \pmod{m}.
\end{equation}
Thus, the support sets for the logical states are related by bitwise complementation:
\[
S_1 = \{\bar{s} : s \in S_0\},\quad\text{where}\quad \bar{s}\text{ denotes the bitwise complement of } s.
\]
Consequently, the logical states can be explicitly expressed as
\[
\ket{0_L}=\sum_{s\in S_0} c_s\ket{s},\quad\text{and}\quad \ket{1_L}=\sum_{s\in S_0} c_s\ket{\bar s},
\]
noting the shared amplitudes $c_s$ due to the transversal bit-flip symmetry.

This symmetry structure significantly simplifies the linear-programming feasibility step. Specifically, KL conditions involving two-qubit operators $Z_iZ_j$ are automatically satisfied, since these operators commute with $\overline{X}=X^{\otimes n}$, implying $\bra{0_L}Z_i Z_j\ket{0_L} = \bra{1_L}Z_i Z_j\ket{1_L}$. Similarly, single-qubit operators $Z_i$ anticommute with $\overline{X}$, forcing their expectation values to vanish: $\bra{0_L}Z_i\ket{0_L} = -\bra{1_L}Z_i\ket{1_L}=0$. Thus, the original LP reduces to the simpler form:
\[
\sum_{s\in S_0}(-1)^{s_i} x_s = 0,\quad\text{for all } i\in[n],\quad\text{with } x_s = |c_s|^2 \geq 0.
\]

\subsection{\texorpdfstring{$((7,2,3))$}{((7,2,3))} Codes with Transversal Group \texorpdfstring{$\mathrm{BD}_{2m}$}{BD₂ₘ}}\label{subsec:723BD16}

We present examples of \(((7,2,3))\) quantum codes with transversal group \(\mathrm{BD}_{2m}\) discovered through our algorithm. Previously, it was known that the following binary dihedral groups appear in \(((7,2,3))\) codes: \(\mathrm{BD}_4\) and \(\mathrm{BD}_8\) occur in stabilizer codes~\cite{cross2025small}, \(\mathrm{BD}_6\) is a subgroup of the transversal group of the Steane code~\cite{steane1996simple}, and \(\mathrm{BD}_{10}\) appears as a subgroup of the permutation-invariant code constructed in~\cite{kubischta2023family}.

Beyond these known cases, our algorithm finds that all \(\mathrm{BD}_{2m}\) groups with \(2m\) ranging from \(12\) to \(36\) can also be realized in \(((7,2,3))\) codes. We highlight three noteworthy examples here — namely, \(\mathrm{BD}_{16}\) (corresponding to a transversal \(T\) gate), \(\mathrm{BD}_{32}\) (corresponding to a transversal \(\sqrt{T}\) gate), and \(\mathrm{BD}_{36}\), which we believe represents the upper bound for realizable \(\mathrm{BD}_{2m}\) transversal groups in \(((7,2,3))\) codes. Other examples are summarized in Appendix~\ref{appendix:723BD2m}.
For \(2m \geq 38\), we have not found any angle vector \(\vec{a}\) that passes the LP filters for both \(\mathrm{BD}_{2m}\) and \(\mathrm{C}_{2m}\).

\subsubsection{\texorpdfstring{$\mathrm{BD}_{16}$}{BD₁₆} (transversal \texorpdfstring{$T$}{T})}

The group \(\mathrm{BD}_{16}\) includes the \(T\) gate as an element. Notably, the smallest stablizer code with transversal $T$ is the $((15,2,3))$ Reed-Muller code~\cite{koutsioumpas2024quantum}, and Kubischta et al.~\cite{kubischta2024permutationinvariantquantumcodestransversal} constructed a $((11,2,3))$ permutation invariant code supporting transversal \(T\) gate. Here, we apply the SS-LP method to construct $((7,2,3))$ codes with transversal \(T\) gate. 

The SS-LP method gives the possible angle vectors $\vec{a}$ as listed below:
\begin{align*}
    \left(1,2,2,2,2,3,3\right),\left(1,1,2,2,3,3,3\right),\left(1,1,2,2,2,3,4\right),\left(1,1,2,2,2,2,5\right),\left(1,1,1,2,3,3,4\right),\left(1,1,1,2,2,4,4\right),\\
    \left(1,1,1,2,2,3,5\right),\left(1,1,1,2,2,2,6\right),\left(1,1,1,1,3,4,4\right),\left(1,1,1,1,3,3,5\right),\left(0,1,2,2,3,3,4\right),\left(0,1,1,2,3,3,5\right).
\end{align*}

Numerical results indicate that
the corresponding range is $\lambda^{*}\in[0.77,2.2]$. 
That is, any $\lambda^{*}$ in this range labels a different $((7,2,3))$ code supporting transversal $T$ gate.

Here we present an example of these codes corresponding to $\vec{a}=\left(1,2,2,2,2,3,3\right)$ and $\lambda^{*}=\sqrt{\tfrac{21}{8}}$.

The logical states are given by:
\begin{align*}
    4\left|0_{L}\right\rangle =&e^{i\theta_{1}}\left|0000000\right\rangle +\sqrt{3}\left|0111100\right\rangle +\sqrt{3}\left|1001110\right\rangle +\sqrt{2}\left|1010101\right\rangle \\
    &+\sqrt{2}\left|1011001\right\rangle +e^{i\theta_{2}}\left|1110010\right\rangle +2i\left|0100011\right\rangle \\
    \left|1_{L}\right\rangle =&X^{\otimes7}\left|0_{L}\right\rangle 
\end{align*}
where \(\theta_1, \theta_2 \in \mathbb{R}\) are free parameters. The transversal gates for this code are:
\[ \overline{X}=X^{\otimes7}, \]
\[ \overline{Z\left(-\tfrac{2\pi}{8}\right)}=Z\left(\tfrac{2\pi}{8}\right)\otimes Z\left(\tfrac{4\pi}{8}\right)\otimes Z\left(\tfrac{4\pi}{8}\right)\otimes Z\left(\tfrac{4\pi}{8}\right)\otimes Z\left(\tfrac{4\pi}{8}\right)\otimes Z\left(\tfrac{6\pi}{8}\right)\otimes Z\left(\tfrac{6\pi}{8}\right). \]
The weight enumerator of this code is:
\[ A=\left[1,0,\frac{21}{8},\frac{2}{8}-\frac{1}{8}a,\frac{215}{16}+\frac{18}{16}a+\frac{1}{16}b,\frac{87}{16}-\frac{42}{16}a-\frac{3}{16}b,\frac{635}{16}+\frac{38}{16}a+\frac{3}{16}b,\frac{25}{16}-\frac{12}{16}a-\frac{1}{16}b\right], \]
\[ B=\left[1,0,\frac{21}{8},\frac{441}{16}+\frac{10}{16}a+\frac{1}{16}b,\frac{352}{16}-3a-\frac{1}{4}b,\frac{1590}{16}+\frac{84}{16}a+\frac{6}{16}b,\frac{846}{16}-4a-\frac{1}{4}b,\frac{809}{16}+\frac{18}{16}a+\frac{1}{16}b\right], \]
with parameters \(a = \cos(2\theta_1) + \cos(2\theta_2)\) and \(b = \cos(2\theta_1 - 2\theta_2)\).

\subsubsection{\texorpdfstring{$\mathrm{BD}_{32}$}{BD₃₂}}

The group \(\mathrm{BD}_{32}\) contains the \(\sqrt{T}\) gate. While~\cite{kubischta2024permutationinvariantquantumcodestransversal} constructed such a transversal gate using $19$ qubits, we construct $((7,2,3))$ codes with transversal \(\sqrt{T}\). He we present one example of such codes.
The logical states are:
\begin{align*}
    \sqrt{32}\left|0_{L}\right\rangle =&\left|0000000\right\rangle +is_{2}\sqrt{1+32\theta^{2}}\left|0001101\right\rangle +is_{3}\sqrt{6-32\theta^{2}}\left|0010101\right\rangle +2is_{4}\left|0100011\right\rangle \\
    &+s_{5}\sqrt{3}\left|0111100\right\rangle +s_{6}\sqrt{7}\left|1011010\right\rangle +s_{7}\sqrt{5}\left|1100110\right\rangle +s_{8}\sqrt{4-32\theta^{2}}\left|1101001\right\rangle +s_{9}\theta\sqrt{32}\left|1110001\right\rangle \\    
    \left|1_{L}\right\rangle =&X^{\otimes7}\left|0_{L}\right\rangle 
\end{align*}
where \(s_i = \pm 1\) and \(\theta \in [0, \sqrt{\tfrac{1}{8}}]\). The associated transversal gates are:
\[ \overline{X}=X^{\otimes7}, \]
\[ \overline{Z\left(-\tfrac{2\pi}{16}\right)}=Z\left(\tfrac{2\pi}{16}\right)\otimes Z\left(\tfrac{3\pi}{16}\right)\otimes Z\left(\tfrac{4\pi}{16}\right)\otimes Z\left(\tfrac{4\pi}{16}\right)\otimes Z\left(\tfrac{5\pi}{16}\right)\otimes Z\left(\tfrac{6\pi}{16}\right)\otimes Z\left(\tfrac{7\pi}{16}\right). \]
The weight enumerator is given by Equation~\eqref{eq:723shorwt-line}, with the signature norm computed as:
\[ \lambda^{*2}=\frac{67}{32}-10\theta^{2}+80\theta^{4}+16s_{2}s_{3}s_{8}s_{9}\theta\sqrt{\left(\frac{1}{16}+\theta^{2}\right)\left(\frac{3}{16}-\theta^{2}\right)\left(\frac{1}{8}-\theta^{2}\right)}. \]

\subsubsection{\texorpdfstring{$\mathrm{BD}_{36}$}{BD₃₆}}

Finally, we present a code realizing transversal group \(\mathrm{BD}_{36}\), which is likely the maximal \(\mathrm{BD}_{2m}\) group achievable within the \(((7,2,3))\) parameter range. The logical states are:
\begin{align*}
    6\left|0_{L}\right\rangle =&e^{i\theta}\left|0000000\right\rangle +\sqrt{2}s_{2}\left|0001110\right\rangle +i\sqrt{3}s_{3}\left|0111100\right\rangle +\sqrt{4}s_{4}\left|0100011\right\rangle \\
    &+i\sqrt{5}s_{5}\left|1100110\right\rangle +i\sqrt{6}s_{6}\left|1101001\right\rangle +i\sqrt{7}s_{7}\left|1011010\right\rangle +\sqrt{8}s_{8}\left|0010101\right\rangle ,\\
    \left|1_{L}\right\rangle =&X^{\otimes7}\left|0_{L}\right\rangle,
\end{align*}
where \(s_i = \pm 1\) and \(\theta \in \mathbb{R}\). The transversal group is implemented via:
\[ \overline{X}=X^{\otimes7} \]
\[ \overline{Z\left(-\tfrac{\pi}{9}\right)}=Z\left(\tfrac{2\pi}{9}\right)\otimes Z\left(\tfrac{3\pi}{9}\right)\otimes Z\left(\tfrac{4\pi}{9}\right)\otimes Z\left(\tfrac{5\pi}{9}\right)\otimes Z\left(\tfrac{6\pi}{9}\right)\otimes Z\left(\tfrac{7\pi}{9}\right)\otimes Z\left(\tfrac{8\pi}{9}\right). \]
The corresponding weight enumerator is:
\[ A=\left[1,0,\frac{161}{81},\frac{7}{81},\frac{1330}{81},\frac{35}{27},\frac{3472}{81},\frac{28}{81}\right]+\cos(2\theta)\cdot\left[0,0,0,\frac{7}{81},-\frac{49}{81},\frac{35}{27},-\frac{91}{81},\frac{28}{81}\right], \]
\[ B=\left[1,0,\frac{161}{81},\frac{721}{27},\frac{497}{27},\frac{3010}{27},\frac{3731}{81},\frac{4079}{81}\right]+\cos(2\theta)\cdot\left[0,0,0,-\frac{7}{27},\frac{112}{81},-\frac{70}{27},\frac{56}{27},-\frac{49}{81}\right].\]

\subsection{\texorpdfstring{$((8,2,3))$}{((8,2,3))} Codes with Transversal Group \texorpdfstring{$\mathrm{BD}_{2m}$}{BD₂ₘ}}

Although our search over \(((7,2,3))\) codes did not yield transversal \(\mathrm{BD}_{2m}\) examples for \(2m > 36\), extending the same framework to \(n=8\) enables the construction of \(((8,2,3))\) codes supporting transversal groups \(\mathrm{BD}_{2m}\) with \(2m>36\). In this subsection, we present explicit examples, including codes realizing transversal gates corresponding to \(\mathrm{BD}_{64}\) (associated with a \(T^{1/4}\) gate) and \(\mathrm{BD}_{84}\). Additional examples with transversal groups \(\mathrm{BD}_{2m}\) for \(2m \in \{38,72,74,76,78,80\}\) are provided in Appendix~\ref{appendix:823-BD2m}.  
For \(2m = 82\) and $2m \ge 86$, we have not found any angle vector \(\vec{a}\) that passes the LP filters for both \(\mathrm{BD}_{2m}\) and \(\mathrm{C}_{2m}\).

\subsubsection{\texorpdfstring{$\mathrm{BD}_{64}$}{BD₆₄}}

An example of a \(((8,2,3))\) code supporting transversal group \(\mathrm{BD}_{64}\) is constructed as follows. The logical basis states are:
\begin{align*}
    8\left|0_{L}\right\rangle =&\sqrt{10}\left|00000000\right\rangle +e^{i\theta_{1}}\left|00001110\right\rangle +is_{1}\sqrt{6}\left|00100011\right\rangle +2is_{2}\left|01010111\right\rangle +\sqrt{8}s_{3}\left|01100110\right\rangle \\
    &+\sqrt{3}e^{i\theta_{2}}\left|01101001\right\rangle +\sqrt{15}is_{4}\left|10111101\right\rangle +2e^{i\theta_{3}}\left|11000101\right\rangle +\sqrt{13}s_{5}\left|11011010\right\rangle,\\
    \left|1_{L}\right\rangle =&X^{\otimes8}\left|0_{L}\right\rangle.
\end{align*}
where \(s_i = \pm 1\) and \(\theta_i \in \mathbb{R}\) are free parameters. The set of transversal gates for this code includes \(\overline{X} = X^{\otimes 8}\) and:
\[ \overline{Z\left(-\tfrac{2\pi}{32}\right)}=Z\left(\tfrac{8\pi}{32}\right)\otimes Z\left(\tfrac{14\pi}{32}\right)\otimes Z\left(\tfrac{20\pi}{32}\right)\otimes Z\left(\tfrac{24\pi}{32}\right)\otimes Z\left(\tfrac{34\pi}{32}\right)\otimes Z\left(\tfrac{46\pi}{32}\right)\otimes Z\left(\tfrac{48\pi}{32}\right)\otimes Z\left(\tfrac{60\pi}{32}\right).\]

\subsubsection{\texorpdfstring{$\mathrm{BD}_{84}$}{BD₈₄}}

We also construct a \(((8,2,3))\) code with transversal group \(\mathrm{BD}_{84}\). The logical basis states are given by:
\begin{align*}
    \sqrt{84}\left|0_{L}\right\rangle =&\sqrt{10}\left|00000000\right\rangle +2e^{i\theta_{1}}\left|00010101\right\rangle +\sqrt{2}e^{i\theta_{2}}\left|00111011\right\rangle +\sqrt{11}e^{i\theta_{3}}\left|01011010\right\rangle +\sqrt{12}e^{i\theta_{4}}\left|01100110\right\rangle \\
    &+\sqrt{3}e^{i\theta_{5}}\left|01101001\right\rangle +\sqrt{17}e^{i\theta_{6}}\left|10001111\right\rangle +3e^{i\theta_{7}}\left|10111100\right\rangle +4ise^{i\theta_{6}}\left|11110001\right\rangle \\
    \left|1_{L}\right\rangle =&X^{\otimes8}\left|0_{L}\right\rangle 
\end{align*}
where \(s_i = \pm 1\) and \(\theta_i \in \mathbb{R}\) are free parameters. Logical \(\ket{1_L}\) is obtained by applying the transversal operator \(\overline{X} = X^{\otimes 8}\) to \(\ket{0_L}\). An additional transversal \(Z\)-axis rotation is implemented as:
\[ \overline{Z\left(-\tfrac{2\pi}{42}\right)}=Z\left(\tfrac{8\pi}{42}\right)\otimes Z\left(\tfrac{20\pi}{42}\right)\otimes Z\left(\tfrac{26\pi}{42}\right)\otimes Z\left(\tfrac{36\pi}{42}\right)\otimes Z\left(\tfrac{44\pi}{42}\right)\otimes Z\left(\tfrac{54\pi}{42}\right)\otimes Z\left(\tfrac{68\pi}{42}\right)\otimes Z\left(\tfrac{78\pi}{42}\right) \]

\section{Conclusion}

In this work, we developed a systematic framework for discovering QECCs with prescribed transversal gate groups, with a particular emphasis on nonadditive constructions beyond the stabilizer paradigm. By combining Stiefel manifold parameterizations with composite loss functions enforcing both the Knill–Laflamme conditions and target transversal symmetries, we uncovered a broad landscape of new $((6,2,3))$, $((7,2,3))$, and $((8,2,3))$ codes admitting various finite subgroups of $\mathrm{SU}(2)$ as their transversal logical operations.

To efficiently search for codes with transversal diagonal gates, we introduced the SS-LP method, which dramatically shrinks the search space by encoding the problem into integer partitions subject to linear constraints. Applying this approach, we systematically explored codes with transversal cyclic groups $\mathrm{C}_{2m}$ and binary dihedral groups $\mathrm{BD}_{2m}$, identifying the smallest known codes supporting transversal $T$, $\sqrt{T}$, and $T^{1/4}$ gates. We also discovered new codes realizing exceptional groups such as the binary icosahedral group $2I$.

These results reveal a far richer structure of nonadditive codes and transversal gate groups than previously recognized, highlighting deep connections between quantum error correction, finite group theory, and the geometry of code spaces.

Looking ahead, several promising directions remain open. Extending our methodology to find codes with higher code dimensions ($K>2$), capable of encoding multiple logical qubits, is an important next step. Similarly, searching for codes with larger minimum distance ($d>3$) could further enhance fault-tolerance properties. More broadly, understanding how transversal and near-transversal operations behave in larger, higher-distance, or multi-qubit codes remains a key challenge for realizing scalable fault-tolerant quantum computation.

\appendix

\section{Finite Subgroups of \texorpdfstring{$\mathrm{SU}(2)$}{SU(2)}}\label{appendix:irrep-su2-subgroup}

In this appendix, we list the inequivalent irreducible representations (irreps) of finite subgroups of \(\mathrm{SU}(2)\), which are used to identify the transversal groups of newly discovered quantum error-correcting codes (QECCs). The irreducible representations are obtained using the method described in~\cite{dixon1970computing}.

\paragraph{Cyclic Groups \texorpdfstring{$\mathrm{C}_{2m}$}{C₂ₘ}.}The cyclic groups \(\mathrm{C}_{2m}\) are Abelian. In particular, \(\mathrm{C}_2 = \{I_2, -I_2\}\) represents the trivial transversal group.

\paragraph{Binary Dihedral Groups \texorpdfstring{$\mathrm{BD}_{2m}$}{BD₂ₘ}.}The binary dihedral groups \(\mathrm{BD}_{2m}\) are non-Abelian. The dimensions of all their irreducible representations are:
\[[1,1,1,1,2,2,\ldots,2],\]
such that the group order satisfies
\[4m = 1 \times 4 + 2^2 \times (m-1).\]
The number of conjugacy classes is distributed as:
\[[1,1,2,2,\ldots,2,m,m],\]
ensuring the total group order is
\[4m = 1 \times 2 + 2 \times (m-1) + m \times 2.\]

Character tables for the exceptional finite subgroups \(2\mathrm{T}\), \(2\mathrm{O}\), and \(2\mathrm{I}\) are given in Tables~\ref{tab:2T-char},~\ref{tab:2O-char}, and~\ref{tab:2I-char}, respectively.

\begin{table}[h]
    \centering
    \renewcommand{\arraystretch}{1.2}
    \begin{tabular}{c|ccccccc}
        \hline
        size & 1 & 1 & 4 & 4 & 4 & 4 & 6 \\
        \hline
        $A_{0}$ & 1 & 1 & 1 & 1 & 1 & 1 & 1 \\
        $A_{1}$ & 1 & 1 & $-a$ & $-a$ & $-a^*$ & $-a^*$ & 1 \\
        $A_{2}$ & 1 & 1 & $-a^*$ & $-a^*$ & $-a$ & $-a$ & 1 \\
        $A_{3}$ & 2 & -2 & $-a$ & $a$ & $a^*$ & $-a^*$ & 0 \\
        $A_{4}$ & 2 & -2 & $-a^*$ & $a^*$ & $a$ & $-a$ & 0 \\
        $A_{5}$ & 2 & -2 & 1 & -1 & -1 & 1 & 0 \\
        $A_{6}$ & 3 & 3 & 0 & 0 & 0 & 0 & -1 \\
        \hline
    \end{tabular}
    \caption{\label{tab:2T-char}Character table of \(2\mathrm{T}\), where \(a = e^{i\pi/3}\).}
\end{table}

\begin{table}[h]
    \centering
    \renewcommand{\arraystretch}{1.2}
    \begin{tabular}{c|cccccccc}
        \hline
        size & 1 & 1 & 6 & 6 & 6 & 8 & 8 & 12 \\
        \hline
        $A_{0}$ & 1 & 1 & 1 & 1 & 1 & 1 & 1 & 1 \\
        $A_{1}$ & 1 & 1 & -1 & -1 & 1 & 1 & 1 & -1 \\
        $A_{2}$ & 2 & -2 & $-\sqrt{2}$ & $\sqrt{2}$ & 0 & 1 & -1 & 0 \\
        $A_{3}$ & 2 & 2 & 0 & 0 & 2 & -1 & -1 & 0 \\
        $A_{4}$ & 2 & -2 & $\sqrt{2}$ & $-\sqrt{2}$ & 0 & 1 & -1 & 0 \\
        $A_{5}$ & 3 & 3 & -1 & -1 & -1 & 0 & 0 & 1 \\
        $A_{6}$ & 3 & 3 & 1 & 1 & -1 & 0 & 0 & -1 \\
        $A_{7}$ & 4 & -4 & 0 & 0 & 0 & -1 & 1 & 0 \\
        \hline
    \end{tabular}
    \caption{\label{tab:2O-char}Character table of \(2\mathrm{O}\).}
\end{table}

\begin{table}[h]
    \centering
    \renewcommand{\arraystretch}{1.2}
    \begin{tabular}{c|ccccccccc}
        \hline
        size & 1 & 1 & 12 & 12 & 12 & 12 & 20 & 20 & 30 \\
        \hline
        $A_{0}$ & 1 & 1 & 1 & 1 & 1 & 1 & 1 & 1 & 1 \\
        $A_{1}$ & 2 & -2 & $a$ & $a^{-1}$ & $-a$ & $-a^{-1}$ & 1 & -1 & 0 \\
        $A_{2}$ & 2 & -2 & $-a^{-1}$ & $-a$ & $a^{-1}$ & $a$ & 1 & -1 & 0 \\
        $A_{3}$ & 3 & 3 & $a$ & $-a^{-1}$ & $a$ & $-a^{-1}$ & 0 & 0 & -1 \\
        $A_{4}$ & 3 & 3 & $-a^{-1}$ & $a$ & $-a^{-1}$ & $a$ & 0 & 0 & -1 \\
        $A_{5}$ & 4 & -4 & 1 & -1 & -1 & 1 & -1 & 1 & 0 \\
        $A_{6}$ & 4 & 4 & -1 & -1 & -1 & -1 & 1 & 1 & 0 \\
        $A_{7}$ & 5 & 5 & 0 & 0 & 0 & 0 & -1 & -1 & 1 \\
        $A_{8}$ & 6 & -6 & -1 & 1 & 1 & -1 & 0 & 0 & 0 \\
        \hline
    \end{tabular}
    \caption{\label{tab:2I-char}Character table of \(2\mathrm{I}\), where \(a = (\sqrt{5}+1)/2\).}
\end{table}

\section{Transversal Group of Known \texorpdfstring{$((6,2,3))$}{((6,2,3))} Codes}\label{appendix:623families}

\subsection{\texorpdfstring{$((6,2,3))$}{((6,2,3))} Codes Built from the \texorpdfstring{$((5,2,3))$}{((5,2,3))} Code}

In previous work~\cite{Cao_2022}, it was demonstrated that extending the number of qubits and applying non-local unitaries can yield new quantum codes inherited from known ones. Using this construction principle, a family of \(((6,2,3))\) codes can be obtained by extending the unique \(((5,2,3))\) stabilizer code. We begin with the logical basis of the \(((5,2,3))\) code, given by
\begin{align*}
    4\left|0^{((5,2,3))}_{L}\right\rangle  =&\left|00000\right\rangle -\left|00011\right\rangle +\left|00101\right\rangle -\left|00110\right\rangle +\left|01001\right\rangle +\left|01010\right\rangle -\left|01100\right\rangle -\left|01111\right\rangle \\&-\left|10001\right\rangle +\left|10010\right\rangle +\left|10100\right\rangle -\left|10111\right\rangle -\left|11000\right\rangle -\left|11011\right\rangle -\left|11101\right\rangle -\left|11110\right\rangle,\\
    4\left|1^{((5,2,3))}_{L}\right\rangle  =&\left|00001\right\rangle +\left|00010\right\rangle +\left|00100\right\rangle +\left|00111\right\rangle +\left|01000\right\rangle -\left|01011\right\rangle -\left|01101\right\rangle +\left|01110\right\rangle \\&+\left|10000\right\rangle +\left|10011\right\rangle -\left|10101\right\rangle -\left|10110\right\rangle +\left|11001\right\rangle -\left|11010\right\rangle +\left|11100\right\rangle -\left|11111\right\rangle .
\end{align*}
This code admits a transversal representation of the binary tetrahedral group \(2\mathrm{T}\), with logical operators \(\overline{X} = X^{\otimes 5}\), \(\overline{Z} = Z^{\otimes 5}\), and \(\overline{F} = (YHS)^{\otimes 5}\) \cite{kubischta2023family}.

To construct a corresponding \(((6,2,3))\) code, we append an ancillary qubit initialized in the \(\ket{0}\) state and apply an arbitrary two-qubit unitary \(U \in \mathrm{SU}(4)\) to the last two qubits:
\begin{equation}\label{eq:623-from-523}
\begin{aligned}
    \left|0^{((6,2,3))}_{L}\right\rangle &= \left(I^{\otimes 4} \otimes U\right)\left(\left|0^{((5,2,3))}_{L}\right\rangle \otimes \ket{0} \right), \\
    \left|1^{((6,2,3))}_{L}\right\rangle &= \left(I^{\otimes 4} \otimes U\right)\left(\left|1^{((5,2,3))}_{L}\right\rangle \otimes \ket{0} \right).
\end{aligned}
\end{equation}
The resulting code family exhibits a quantum weight enumerator parameterized as in Equation~\eqref{eq:623-shorwt-SU4}.

Regardless of the choice of two-qubit gate \(U\), at least two logical gates remain transversal:
\[ \overline{X}=X\otimes I\otimes Y\otimes Y\otimes I\otimes I,\quad \overline{Z}=Y\otimes Z\otimes Y\otimes I\otimes I\otimes I,\]
inherited from the first four qubits of the \(((5,2,3))\) code. Consequently, the transversal group of codes constructed via Equation~\eqref{eq:623-from-523} must be either \(\mathrm{BD}_4\) or \(2\mathrm{T}\), depending on the specific form of \(U\). We now present explicit constructions for each case. For the \(\mathrm{BD}_4\) group, one may choose
\[ U=\mathrm{diag}\left(1,1,1,-1\right)\left(I_{2}\otimes R_{y}\left(\theta\right)\right), \]
with \(\cos\theta \neq 1\). In this case, no additional independent transversal operator beyond \(\overline{X}\) and \(\overline{Z}\) can be found. To realize the full \(2\mathrm{T}\) group, a suitable \(U\) can be
\[ U=\mathrm{diag}\left(1,1,1,-1\right)\left(\left( \sqrt{\frac{3-\sqrt{3}}{12}} (X+Y)- \sqrt{\frac{3+\sqrt{3}}{6}}Z\right)\otimes R_{y}\left(\theta\right)\right), \]
which allows for an additional transversal logical gate
\[\overline{F} = (YHS)^{\otimes 4} \otimes Z\left(\tfrac{2\pi}{3}\right) \otimes I.\]
Together, \(\overline{X}\), \(\overline{Z}\), and \(\overline{F}\) generate the \(2\mathrm{T}\) transversal group. For both constructions above, the corresponding weight enumerators take the form of Equation~\eqref{eq:623-shorwt-SU4} with parameter \(a = (\cos\theta)^2\).

\subsection{\texorpdfstring{$((6,2,3))$}{((6,2,3))} \texorpdfstring{$\mathrm{SO}(5)$}{SO(5)} Code Family}

A distinct family of \(((6,2,3))\) codes was introduced in~\cite{du2024characterizing} by searching over a continuous range of the parameter \(\lambda^*\). These codes are inequivalent to the \(((6,2,3))\) codes derived from the \(((5,2,3))\) stabilizer code discussed previously. The construction is parametrized by a special orthogonal matrix \(W \in \mathrm{SO}(5)\), with the logical basis of the code defined as follows.

Let \(\left\{\ket{S_j}\right\}_{j=1}^5\) be the entangled state components:
\[ \sqrt{2}\left|S_{1}\right\rangle =\left|00001\right\rangle +\left|11110\right\rangle ,\sqrt{2}\left|S_{2}\right\rangle =\left|00010\right\rangle +\left|11101\right\rangle ,\sqrt{2}\left|S_{3}\right\rangle =\left|00100\right\rangle +\left|11011\right\rangle \]
\[ \sqrt{2}\left|S_{4}\right\rangle =\left|01000\right\rangle +\left|10111\right\rangle ,\sqrt{2}\left|S_{5}\right\rangle =\left|10000\right\rangle +\left|01111\right\rangle \]
Then the logical basis states are defined as:
\begin{equation}
\begin{aligned}
    \left|0_{L}\right\rangle =&\frac{1}{2}\sum_{j=1}^{5}\left(\left(W_{j,1}+iW_{j,2}\right)\left|0\right\rangle +\left(W_{j,3}+iW_{j,4}\right)\left|1\right\rangle \right)\otimes\left|S_{j}\right\rangle \\
    \left|1_{L}\right\rangle =&\frac{1}{2}\sum_{j=1}^{5}\left(\left(W_{j,3}-iW_{j,4}\right)\left|0\right\rangle +\left(iW_{j,2}-W_{j,1}\right)\left|1\right\rangle \right)\otimes\left|S_{j}\right\rangle 
\end{aligned}
\end{equation}
The quantum weight enumerator for this code family is given by Equation~\eqref{eq:623-shorwt-SO5}, where the parameter \(a\) is computed as \(a = \sum_{j=1}^5 W_{j,5}^4\). Notably, the weight enumerator depends solely on the fifth column of the matrix \(W\). Moreover, it has been proven~\cite{du2024characterizing} that all such codes with identical last columns are locally unitary (LU) equivalent. The remaining columns of \(W\) determine only the local unitary applied to the first qubit and a rotation within the logical subspace.

When \(a = \sum_{j=1}^5 W_{j,5}^4 \ne 1\), the resulting codes are non-stabilizer, as indicated by the non-integer entries in their weight enumerators. These codes also exhibit transversal gate structures distinct from those of stabilizer codes. For instance, any stabilizer code must support \(\mathrm{BD}_4\) as a subgroup of its transversal group, due to the required presence of transversal logical \(X\) and \(Z\) gates. In contrast, the \(\mathrm{SO}(5)\) code family supports only limited transversal groups: \(\mathrm{C}_2\), \(\mathrm{C}_4\), or \(\mathrm{BD}_4\), depending on the sparsity pattern of the last column \([W_{1,5}, \dots, W_{5,5}]\) as follows:
\begin{enumerate}
    \item If at most two components are nonzero (e.g., \([W_{1,5}, W_{2,5}, 0, 0, 0]\)), the transversal group is \(\mathrm{BD}_4 \cong \{\pm I, \pm iX, \pm iY, \pm iZ\}\).
    \item If exactly three components are nonzero (e.g., \([W_{1,5}, W_{2,5}, W_{3,5}, 0, 0]\)), the group reduces to \(\mathrm{C}_4 \cong \{\pm I, \pm iZ\}\).
    \item If four or more components are nonzero, the code admits only a trivial transversal group: \(\mathrm{C}_2 \cong \{\pm I\}\).
\end{enumerate}
Further examples corresponding to each case will be presented in the following paragraphs.

\paragraph{\texorpdfstring{$((6,2,3))$}{((6,2,3))} \texorpdfstring{$\mathrm{SO}(5)$}{SO(5)} Code with Transversal Group \texorpdfstring{$\mathrm{BD}_4$}{BD₄}}If the last column of the matrix \(W\), namely \([W_{1,5}, W_{2,5}, W_{3,5}, W_{4,5}, W_{5,5}]\), contains at most two nonzero elements, we may—without loss of generality—reorder the qubits such that only the first two entries are nonzero. In this case, we can choose a canonical representative of the family with matrix \(W\) in the following form:
\[ W=\begin{bmatrix}
-W_{2,5} & W_{2,5} & -W_{2,5} & W_{2,5} & W_{1,5}\\
W_{1,5} & -W_{1,5} & W_{1,5} & -W_{1,5} & W_{2,5}\\
1/2 & 1/2 & 1/2 & 1/2 & 0\\
1/2 & -1/2 & -1/2 & 1/2 & 0\\
1/2 & 1/2 & -1/2 & -1/2 & 0
\end{bmatrix}. \]
For codes constructed from this form, we identify the following two transversal logical operators:
\[ \overline{X}=Y\otimes\left(\tfrac{\sqrt{3}X+Y}{2}\right)\otimes\left(\tfrac{X-\sqrt{3}Y}{2}\right)\otimes\left(\tfrac{\sqrt{3}X+Y}{2}\right)\otimes\left(\tfrac{X-\sqrt{3}Y}{2}\right)\otimes\left(\tfrac{X-\sqrt{3}Y}{2}\right), \]
\[ \overline{Z}=X\otimes Z\otimes Z\otimes I\otimes I\otimes I.\]
These codes admit a weight enumerator of the form given in Equation~\eqref{eq:623-shorwt-SO5}, with parameter \(a \in \left[\tfrac{1}{2}, 1\right]\). This regime characterizes a subclass of \(\mathrm{SO}(5)\) codes with transversal group \(\mathrm{BD}_4\).

\paragraph{\texorpdfstring{$((6,2,3))$}{((6,2,3))} \texorpdfstring{$\mathrm{SO}(5)$}{SO(5)} Code with Transversal Group \texorpdfstring{$\mathrm{C}_4$}{C₄}}For cases where the last column of \(W\) contains exactly three nonzero entries (e.g., \([W_{1,5}, W_{2,5}, W_{3,5}, 0, 0]\)), the matrix \(W\) can be parameterized in the following structured form:
\[ W=\begin{bmatrix}
W_{1,1} & W_{1,2} & W_{1,1} & W_{1,2} & W_{1,5}\\
W_{2,1} & W_{2,2} & W_{2,1} & W_{2,2} & W_{2,5}\\
W_{3,1} & W_{3,1} & W_{3,1} & W_{3,1} & W_{3,5}\\
1/2 & -1/2 & -1/2 & 1/2 & 0\\
1/2 & 1/2 & -1/2 & -1/2 & 0
\end{bmatrix},\]
where the entries satisfy
\[ \begin{cases}
W_{1,1}=\frac{1}{2}\sqrt{W_{3,5}^{2}+1}\cos\alpha\\
W_{2,1}=\frac{1}{2}\sqrt{W_{3,5}^{2}+1}\sin\alpha\\
W_{1,2}=\frac{1}{2}\sqrt{W_{3,5}^{2}+1}\cos\beta\\
W_{2,2}=\frac{1}{2}\sqrt{W_{3,5}^{2}+1}\sin\beta\\
W_{3,1}=-\frac{W_{1,1}W_{1,5}+W_{2,1}W_{2,5}}{W_{3,5}}
\end{cases} \]
and
\[ \alpha=\arccos\left(\tfrac{W_{3,5}}{\sqrt{W_{3,5}^{2}+1}}\right)+\arctan\tfrac{W_{2,5}}{W_{1,5}}, \]
\[ \beta=-\arccos\left(\tfrac{W_{3,5}}{\sqrt{W_{3,5}^{2}+1}}\right)+\arctan\tfrac{W_{2,5}}{W_{1,5}}. \]
As \(W_{3,5} \to 0\), this construction asymptotically recovers the \(\mathrm{BD}_4\) case. However, for nonzero \(W_{3,5}\), the only transversal logical operator we find is $\overline{Z}=X\otimes Z\otimes Z\otimes I\otimes I\otimes I$, implying that the transversal group is limited to \(\mathrm{C}_4\). The associated weight enumerator again follows Equation~\eqref{eq:623-shorwt-SO5}, with the parameter range \(a \in \left[\tfrac{1}{3}, 1\right)\).

\paragraph{\texorpdfstring{$((6,2,3))$}{((6,2,3))} \texorpdfstring{$\mathrm{SO}(5)$}{SO(5)} Code with Transversal Group \texorpdfstring{$\mathrm{C}_2$}{C₂}}In the generic case, where the fifth column of \(W\) has four or more nonzero components (i.e., no more than one zero entry), our optimization algorithm fails to identify any nontrivial transversal logical operator beyond the identity. This strongly suggests that the transversal group is trivial, namely \(\mathrm{C}_2 \cong \{\pm I\}\), a scenario impossible for stabilizer codes. These codes still follow the weight enumerator form in Equation~\eqref{eq:623-shorwt-SO5}, but with the parameter \(a\) ranging over \(\left[\tfrac{1}{5}, 1\right)\).

\section{More \texorpdfstring{$((7,2,3))$}{((7,2,3))} Codes with Exceptional Transversal Groups}\label{appendix:723exceptional}

\subsection{\texorpdfstring{$((7,2,3))$}{((7,2,3))} Code with Transversal Group \texorpdfstring{$2\mathrm{T}$}{2T}}

In~\cite{du2024characterizing}, a family of \(((7,2,3))\) codes with cyclic symmetry was introduced, though their transversal properties were not analyzed in detail. Here, we demonstrate that this family can exhibit transversal gate groups isomorphic to the binary tetrahedral group \(2\mathrm{T}\), and in special cases, even extend to \(2\mathrm{O}\) or \(2\mathrm{I}\).

The logical basis states of this code family are defined as:
\begin{equation}\label{eq:723cyclic}
\begin{aligned}
    \left|0_{L}\right\rangle =&c_{0}\left|0000000\right\rangle +\frac{c_{1}}{2\sqrt{3}}\left(\left|\{0000011\}\right\rangle +\left|\{0000101\}\right\rangle +\left|\{0001001\}\right\rangle -3\left|\{0111111\}\right\rangle \right)\\&+c_{2}\left|\{0010111\}\right\rangle +\frac{c_{3}}{2}\left(\left|\{0001111\}\right\rangle +\left|\{0011011\}\right\rangle +\left|\{0011101\}\right\rangle +\left|\{0101011\}\right\rangle \right),\\
    \left|1_{L}\right\rangle=&X^{\otimes 7}\left|0_{L}\right\rangle,
\end{aligned}
\end{equation}
where \(\ket{\{x\}}\) denotes the cyclic orbit of a bit string \(\ket{x}\), i.e.,
\[ \left|\{0000101\}\right\rangle =\left|0000101\right\rangle +\left|1000010\right\rangle +\left|0100001\right\rangle +\left|1010000\right\rangle +\left|0101000\right\rangle +\left|0010100\right\rangle +\left|0001010\right\rangle .\]
The coefficients \(c_0, c_1, c_2, c_3\) are parametrized by \(\lambda^* \in [0, \sqrt{7}]\) and discrete signs \(s_0, s_1 = \pm 1\), satisfying:
\begin{equation}
\begin{cases}
    8c_{0}=\sqrt{\sqrt{7}\lambda^{*}+8},\\
    4c_{1}=s_{0}\sqrt{\sqrt{7}\lambda^{*}},\\
    c_{2}=-2c_{3}+\sqrt{7}c_{0},\\
    \tfrac{5}{2}c_{3}=\sqrt{7}c_{0}+s_{1}\sqrt{7c_{0}^{2}-\frac{15\sqrt{7}\lambda^{*}}{64}}.
\end{cases}
\end{equation}
The code admits a natural implementation of \(2\mathrm{T}\) as its transversal group via:
\[ \overline{X}=X^{\otimes7},\quad \overline{F}=\left(F^{*}\right)^{\otimes7}.\]
At the special point \(\lambda^* = 0\), the code reduces—up to qubit permutation—to the Steane code, which possesses transversal symmetry group \(2\mathrm{O}\). In this case, an additional transversal generator becomes available:
\[ \overline{S}=\left(S^{*}\right)^{\otimes7}, \]
yielding a complete transversal group \(2\mathrm{O} \cong \langle X, F, S \rangle\).

When \(\lambda^* = \sqrt{7}\), the code becomes permutation-invariant and supports a transversal implementation of the binary icosahedral group \(2\mathrm{I}\), consistent with the findings of~\cite{pollatsek2004permutationally}. The exact form of the third transversal generator depends on the sign choice \(s_0\):
\[
\text{If } s_0 = +1: \quad \overline{\Phi^{\star}} = \Phi^{\otimes 7}; \quad
\text{if } s_0 = -1: \quad \overline{\Phi} = (\Phi^{\star})^{\otimes 7}.
\]
In either case, the transversal group is isomorphic to \(2\mathrm{I} \cong \langle X, F, \Phi \rangle\) or \(\langle X, F, \Phi^{\star} \rangle\), respectively. The weight enumerators for this code family vary continuously with \(\lambda^*\) and lie on the parameterized line segment given by Equation~\eqref{eq:723shorwt-line}.

\subsection{\texorpdfstring{$((7,2,3))$}{((7,2,3))} Code with Transversal Group \texorpdfstring{$2\mathrm{O}$}{2O}}

While a \(((7,2,3))\) stabilizer code supporting transversal symmetry group \(2\mathrm{O}\) already exists (notably the Steane code~\cite{steane1996simple}), we present here a new \(((7,2,3))\) code characterized by the unique feature that all its nontrivial transversal operations act exclusively on the first five qubits. 

The code subspace is constructed within a six-dimensional space spanned by states \(\{\ket{S_0}, \ldots, \ket{S_5}\}\), defined as follows:
\begin{align*}
    2\sqrt{3}\left|S_{0}\right\rangle =&\left|0000000\right\rangle +\left|0001111\right\rangle +\left|0010100\right\rangle +\left|1100011\right\rangle +\left|1101100\right\rangle +\left|1110111\right\rangle \\&-\left|0000011\right\rangle -\left|0001100\right\rangle -\left|0010111\right\rangle -\left|1100000\right\rangle -\left|1101111\right\rangle -\left|1110100\right\rangle ,\\
    4\left|S_{1}\right\rangle =&\left|0100100\right\rangle +\left|0101000\right\rangle +\left|0110011\right\rangle +\left|0111100\right\rangle +\left|1000100\right\rangle +\left|1001000\right\rangle +\left|1010011\right\rangle +\left|1011100\right\rangle \\&-\left|0100111\right\rangle -\left|0101011\right\rangle -\left|0110000\right\rangle -\left|0111111\right\rangle -\left|1000111\right\rangle -\left|1001011\right\rangle -\left|1010000\right\rangle -\left|1011111\right\rangle, \\
    2\sqrt{2}\left|S_{2}\right\rangle =&\left|0100000\right\rangle +\left|0100011\right\rangle +\left|1010100\right\rangle +\left|1010111\right\rangle -\left|0110100\right\rangle -\left|0110111\right\rangle -\left|1000000\right\rangle -\left|1000011\right\rangle ,\\
    2\sqrt{2}\left|S_{3}\right\rangle =&\left|0100001\right\rangle +\left|0110101\right\rangle +\left|1000010\right\rangle +\left|1010110\right\rangle -\left|0100010\right\rangle -\left|0110110\right\rangle -\left|1000001\right\rangle -\left|1010101\right\rangle ,\\
    2\left|S_{4}\right\rangle =&\left|0011000\right\rangle +\left|1111011\right\rangle -\left|0011011\right\rangle -\left|1111000\right\rangle ,\\
    2\left|S_{5}\right\rangle =&\left|0101101\right\rangle +\left|1001110\right\rangle -\left|0101110\right\rangle -\left|1001101\right\rangle .
\end{align*}
The logical basis states of the code are given by:
\begin{align*}
    \left|0_{L}\right\rangle =&\frac{1}{4}\left|S_{0}\right\rangle -\frac{1}{2}\left|S_{1}\right\rangle +\frac{1}{2}s_{1}\left|S_{2}\right\rangle +\frac{1}{\sqrt{12}}s_{2}\left|S_{3}\right\rangle -\frac{\sqrt{3}}{4}\left|S_{4}\right\rangle +\frac{1}{\sqrt{6}}s_{2}\left|S_{5}\right\rangle ,\\
    \left|1_{L}\right\rangle =&X^{\otimes5}\otimes I^{\otimes2}\left|0_{L}\right\rangle,
\end{align*}
where \(s_1, s_2 \in \{\pm 1\}\) are discrete parameters. This code supports a transversal implementation of the binary octahedral group \(2\mathrm{O}\), realized via:
\[ \overline{X}=X^{\otimes5}\otimes I^{\otimes2}, \]
\[ \overline{\square}=\left(\frac{iX+i\sqrt{3}Z}{2}\right)^{\otimes2}\otimes\left(\frac{\sqrt{2}I+iX-iZ}{2}\right)\otimes\left(\frac{\sqrt{2}I-iX-iZ}{2}\right)\otimes\left(\frac{\sqrt{2}I+iX+iZ}{2}\right)\otimes I\otimes I,\]
where \(\square = Y(\tfrac{\pi}{4})Z(\tfrac{\pi}{2})Y(-\tfrac{\pi}{4})\). This defines a nonstandard but isomorphic representation of the binary octahedral group:
\[ 2\mathrm{O} \cong \langle H, Z(\tfrac{\pi}{2}) \rangle \cong \langle Y(\tfrac{\pi}{4}) H Y(-\tfrac{\pi}{4}), Y(\tfrac{\pi}{4}) Z(\tfrac{\pi}{2}) Y(-\tfrac{\pi}{4}) \rangle = \langle X, \square \rangle. \]
The quantum weight enumerator of this code lies on the curve defined by Equation~\eqref{eq:723shorwt-line}, corresponding to \(\lambda^* = 1\). Notably, this code differs from all known \(((7,2,3))\) stabilizer codes, either in its transversal gate structure or weight enumerator, despite the fact that its enumerators contain only integers. Therefore, we conclude that this construction defines a non-stabilizer \(((7,2,3))\) code.

\section{More \texorpdfstring{$((7,2,3))$}{((7,2,3))} Codes with Transversal Group \texorpdfstring{$\mathrm{BD}_{2m}$}{BD₂ₘ}}\label{appendix:723BD2m}

\subsection{\texorpdfstring{$\mathrm{BD}_{12}$}{BD₁₂}}

We present two inequivalent \(((7,2,3))\) codes with transversal group \(\mathrm{BD}_{12}\).

\paragraph{Type I.} The logical basis states are:
\begin{align*}
    \sqrt{120}\left|0_{L}\right\rangle =&s_{1}\sqrt{10}\left|0000000\right\rangle +s_{2}\sqrt{12}\left|0110101\right\rangle +s_{3}\sqrt{5}\left|0110110\right\rangle +s_{4}\sqrt{8}\left|0111010\right\rangle +s_{5}\sqrt{5}\left|0111100\right\rangle \\
    &+s_{6}\sqrt{18}\left|1010101\right\rangle -s_{2}s_{4}s_{6}\sqrt{12}\left|1011010\right\rangle +s_{7}\sqrt{15}\left|1100011\right\rangle -s_{3}s_{5}s_{7}\sqrt{15}\left|1101001\right\rangle +is_{8}\sqrt{20}\left|0001110\right\rangle, \\
    \left|1_{L}\right\rangle =&X^{\otimes7}\left|0_{L}\right\rangle,
\end{align*}
where $s_i=\pm 1$. The transversal logical gates are:
\[ \overline{X}=X^{\otimes7}, \]
\[ \overline{Z\left(-\tfrac{2\pi}{6}\right)}=Z\left(\tfrac{2\pi}{6}\right)\otimes Z\left(\tfrac{2\pi}{6}\right)\otimes Z\left(\tfrac{2\pi}{6}\right)\otimes Z\left(\tfrac{4\pi}{6}\right)\otimes Z\left(\tfrac{4\pi}{6}\right)\otimes Z\left(\tfrac{4\pi}{6}\right)\otimes Z\left(\tfrac{4\pi}{6}\right).\]
The weight enumerator follows Equation~\eqref{eq:723shorwt-line} with \(\lambda^{*2} = \frac{269}{150}\).

\paragraph{Type II.} The logical basis states are:
\begin{align*}
    \left|0_{L}\right\rangle =&\frac{\theta}{\sqrt{8}}\left(\left|0000000\right\rangle +\left|1100101\right\rangle +\left|1101001\right\rangle +\left|1110001\right\rangle -\left|0000011\right\rangle -\left|1100110\right\rangle -\left|1101010\right\rangle -\left|1110010\right\rangle \right)\\
    &+is_{1}\sqrt{\frac{1}{12}-\frac{1}{8}\theta^{2}}\left(\left|0100101\right\rangle +\left|0100110\right\rangle +\left|0110001\right\rangle +\left|0101001\right\rangle +\left|0101010\right\rangle +\left|0110010\right\rangle +\left|1000000\right\rangle +\left|1000011\right\rangle \right)\\
    &+\frac{is_{2}\theta}{2}\left(\left|0011100\right\rangle +\left|0011111\right\rangle \right)+\frac{s_{1}s_{2}}{2}\sqrt{\frac{2}{3}-\theta^{2}}\left(\left|1011100\right\rangle -\left|1011111\right\rangle \right)\\
    \left|1_{L}\right\rangle =&X^{\otimes7}\left|0_{L}\right\rangle 
\end{align*}
where \(s_i = \pm 1\) and \(\theta \in \left[0, \sqrt{\tfrac{2}{3}}\right]\). The transversal logical gates are:
\[ \overline{X}=X^{\otimes7}, \]
\[ \overline{Z\left(\tfrac{2\pi}{6}\right)}=Z\left(0\right)\otimes Z\left(\tfrac{2\pi}{6}\right)\otimes Z\left(\tfrac{4\pi}{6}\right)\otimes Z\left(\tfrac{4\pi}{6}\right)\otimes Z\left(\tfrac{4\pi}{6}\right)\otimes Z\left(\pi\right)\otimes Z\left(\pi\right). \]
The weight enumerator is given by Equation~\eqref{eq:723shorwt-line} with the signature norm being
\[ \lambda^{*2}=\frac{29}{9}-8\theta^{2}+12\theta^{4}. \]

\subsection{\texorpdfstring{$\mathrm{BD}_{14}$}{BD₁₄}}

The logical basis states are:
\begin{align*}
    \sqrt{14}\left|0_{L}\right\rangle =&\left|0000000\right\rangle +i\sqrt{2}\left|0001110\right\rangle +\left|1010110\right\rangle +\sqrt{2}\left|1011001\right\rangle +is_{1}\theta\sqrt{14}\left|0010011\right\rangle +is_{2}\sqrt{1-14\theta^{2}}\left|0010101\right\rangle \\
    &+s_{3}\sqrt{\frac{3}{7}+2\theta^{2}}\left(\sqrt{3}\left|0111100\right\rangle +2\left|1100101\right\rangle \right)+s_{4}\sqrt{\frac{4}{7}-2\theta^{2}}\left(\sqrt{3}\left|0111010\right\rangle -2\left|1100011\right\rangle \right)\\
    \left|1_{L}\right\rangle =&X^{\otimes7}\left|0_{L}\right\rangle 
\end{align*}
where $s_i=\pm 1$ and $\theta\in\left[0,\sqrt{\tfrac{1}{14}}\right]$. The transversal gates are:
\[ \overline{X}=X^{\otimes7}, \]
\[ \overline{Z\left(-\tfrac{2\pi}{7}\right)}=Z\left(\tfrac{\pi}{7}\right)\otimes Z\left(\tfrac{2\pi}{7}\right)\otimes Z\left(\tfrac{3\pi}{7}\right)\otimes Z\left(\tfrac{4\pi}{7}\right)\otimes Z\left(\tfrac{5\pi}{7}\right)\otimes Z\left(\tfrac{5\pi}{7}\right)\otimes Z\left(\tfrac{6\pi}{7}\right). \]
The Shor weight enumerator is given by Equation~\eqref{eq:723shorwt-line} with the corresponding signature norm:
\[ \lambda^{*2}=\frac{3701}{2401}-\frac{1384}{343}\theta^{2}+\frac{2768}{49}\theta^{4}+\frac{16s_{1}s_{2}s_{3}s_{4}}{7}\theta\sqrt{\left(\frac{1}{14}-\theta^{2}\right)\left(\frac{3}{14}+\theta^{2}\right)\left(\frac{2}{7}-\theta^{2}\right)}. \]

\subsection{\texorpdfstring{$\mathrm{BD}_{16}$}{BD₁₆}}

In addition to the construction described in Section~\ref{subsec:723BD16}, we present here another inequivalent \(((7,2,3))\) code with transversal group \(\mathrm{BD}_{16}\). The logical basis states are:
\begin{align*}
    4\left|0_{L}\right\rangle =&i\sqrt{3}\left|0000000\right\rangle +i\sqrt{3}\left|0000011\right\rangle +e^{i\theta_{1}}\left|0111101\right\rangle +e^{-i\theta_{1}}\left|0111110\right\rangle +e^{i\theta_{2}}\left|1011101\right\rangle +e^{-i\theta_{2}}\left|1011110\right\rangle \\
    &+e^{i\theta_{3}}\left|1101101\right\rangle +e^{-i\theta_{3}}\left|1101110\right\rangle +e^{i\theta_{4}}\left|1110101\right\rangle +e^{-i\theta_{4}}\left|1110110\right\rangle +e^{i\theta_{5}}\left|1111001\right\rangle +e^{-i\theta_{5}}\left|1111010\right\rangle ,\\
    \left|1_{L}\right\rangle =&X^{\otimes 5} \otimes I^{\otimes 2}\left|0_{L}\right\rangle,
\end{align*}
where $\theta_i\in\mathbb{R}$. The transversal gates are:
\[ \overline{X}=X^{\otimes 5} \otimes I^{\otimes 2}, \]
\[ \overline{Z\left(-\tfrac{\pi}{4}\right)}=Z\left(\tfrac{3\pi}{4}\right)\otimes Z\left(\tfrac{3\pi}{4}\right)\otimes Z\left(\tfrac{3\pi}{4}\right)\otimes Z\left(\tfrac{3\pi}{4}\right)\otimes Z\left(\tfrac{3\pi}{4}\right)\otimes Z\left(\pi\right)\otimes Z\left(-\pi\right). \]
The weight enumerator is described by Equation~\eqref{eq:723shorwt-line}, with signature norm:
\[ \lambda^{*2}=\frac{53}{16}+\frac{1}{16}\sum_{i,j}\cos\left(2\theta_{i}-2\theta_{j}\right). \]

A special case arises when all parameters \(\theta_i\) are set to zero. The logical states then simplify to:
\begin{align*}
    4\left|0_{L}\right\rangle &=i\sqrt{3}\left|0000000\right\rangle +i\sqrt{3}\left|0000011\right\rangle +\left|0111101\right\rangle +\left|0111110\right\rangle +\left|1011101\right\rangle +\left|1011110\right\rangle \\
    &+\left|1101101\right\rangle +\left|1101110\right\rangle +\left|1110101\right\rangle +\left|1110110\right\rangle +\left|1111001\right\rangle +\left|1111010\right\rangle \\
    4\left|1_{L}\right\rangle &=i\sqrt{3}\left|1111111\right\rangle +i\sqrt{3}\left|1111100\right\rangle +\left|1000010\right\rangle +\left|1000001\right\rangle +\left|0100010\right\rangle +\left|0100001\right\rangle \\
    &+\left|0010010\right\rangle +\left|0010001\right\rangle +\left|0001010\right\rangle +\left|0001001\right\rangle +\left|0000110\right\rangle +\left|0000101\right\rangle 
\end{align*}
Although the resulting code is not a stabilizer code (since its weight enumerators are not all integers), we can still identify several stabilizers within the code subspace:
\[ S_{12},S_{23},S_{45},S_{67},X_{6}X_{7} \]
where \(S_{ij}\) denotes the swap operator between qubits \(i\) and \(j\). It is important to note that the logical \(X\) operator implementation differs from that in Section~\ref{subsec:723BD16}, confirming that the two code families are not local-unitary equivalent, even though both belong to the \(\mathrm{BD}_{16}\) class.

\subsection{\texorpdfstring{$\mathrm{BD}_{18}$}{BD₁₈}}

We present two distinct code families realizing transversal group \(\mathrm{BD}_{18}\).

\paragraph{Type I.} The logical basis states are:
\begin{align*}
    \left|0_{L}\right\rangle =&i\sqrt{\frac{1}{18}}\left|0000000\right\rangle \pm i\sqrt{\frac{5}{18}}\left|0000111\right\rangle +\theta_{1}\left|0111001\right\rangle +\theta_{2}\left|0111010\right\rangle +\theta_{3}\left|0111100\right\rangle \\&+\theta_{4}\left|1011001\right\rangle +\theta_{5}\left|1011010\right\rangle +\theta_{6}\left|1011100\right\rangle +\theta_{7}\left|1101001\right\rangle +\theta_{8}\left|1101010\right\rangle \\
    &+\theta_{9}\left|1101100\right\rangle +\theta_{10}\left|1110001\right\rangle +\theta_{11}\left|1110010\right\rangle +\theta_{12}\left|1110100\right\rangle \\
    \left|1_{L}\right\rangle =&X^{\otimes7}\left|0_{L}\right\rangle 
\end{align*}
where the parameters \(\theta_i^2\) satisfy the linear constraint:
\[ \begin{cases}
\theta_{1}^{2}+\theta_{2}^{2}+\theta_{3}^{2}=\frac{1}{6}\\
\theta_{4}^{2}+\theta_{5}^{2}+\theta_{6}^{2}=\frac{1}{6}\\
\theta_{7}^{2}+\theta_{8}^{2}+\theta_{9}^{2}=\frac{1}{6}\\
\theta_{10}^{2}+\theta_{11}^{2}+\theta_{12}^{2}=\frac{1}{6}\\
\theta_{1}^{2}+\theta_{4}^{2}+\theta_{7}^{2}+\theta_{10}^{2}=\frac{2}{9}\\
\theta_{2}^{2}+\theta_{5}^{2}+\theta_{8}^{2}+\theta_{11}^{2}=\frac{2}{9}
\end{cases}. \]
A trivial solution is \(\theta_i^2 = \frac{1}{18}\). Alternatively, an explicit one-parameter family of solutions can be obtained by fixing five independent \(\theta_i^2\) values (e.g., \(\theta_1^2, \theta_2^2, \theta_4^2, \theta_5^2, \theta_7^2\)) to \(\frac{1}{18}\), leaving one degree of freedom for the remaining parameters. The transversal gates are:
\[ \overline{X}=X^{\otimes7}, \]
\[ \overline{Z\left(-\tfrac{2\pi}{9}\right)}=Z\left(\tfrac{4\pi}{9}\right)\otimes Z\left(\tfrac{4\pi}{9}\right)\otimes Z\left(\tfrac{4\pi}{9}\right)\otimes Z\left(\tfrac{4\pi}{9}\right)\otimes Z\left(\tfrac{6\pi}{9}\right)\otimes Z\left(\tfrac{6\pi}{9}\right)\otimes Z\left(\tfrac{6\pi}{9}\right). \]

\paragraph{Type II.} The logical basis states are:
\begin{align*}
    \left|0_{L}\right\rangle =&\frac{e^{i\theta}}{\sqrt{18}}\left|0000000\right\rangle +bs_{1}\left|0111010\right\rangle +s_{2}\sqrt{\frac{1}{18}-b^{2}}\left|0111100\right\rangle +s_{3}d\left|1011010\right\rangle \\
    &-s_{1}s_{2}s_{3}\sqrt{\frac{1}{9}-b^{2}}\left|1100011\right\rangle -s_{3}\sqrt{b^{2}+d^{2}+\frac{1}{18}}\left|1100101\right\rangle +s_{4}\sqrt{\frac{1}{6}-d^{2}}\left(\left|1010110\right\rangle +\left|1101001\right\rangle \right)\\
    &-\frac{s_{1}s_{3}s_{4}}{\sqrt{18}}\left(\sqrt{2}\left|0110110\right\rangle -i\sqrt{2}\left|0001110\right\rangle +i\sqrt{3}\left|0011001\right\rangle \right)\\
    \left|1_{L}\right\rangle =&X^{\otimes7}\left|0_{L}\right\rangle 
\end{align*}
where \(s_i = \pm 1\) and \(\theta \in \mathbb{R}\). Scalars \(b\) and \(d\) are roots of the following polynomials:
\[ \left\{ b\in\left(0,\frac{1}{3}\right):3222b^{2}-49329b^{4}-58320b^{6}+209952b^{8}=25\right\}, \]
\[ d=\frac{1}{\sqrt{36b^{2}+14}}. \]
Numerical approximations yield \(b \approx 0.094956415409\) or \(b \approx 0.230377064\). The transversal group is generated by:
\[ \overline{X}=X^{\otimes7}, \]
\[ \overline{Z\left(-\tfrac{2\pi}{9}\right)}=Z\left(\tfrac{2\pi}{9}\right)\otimes Z\left(\tfrac{2\pi}{9}\right)\otimes Z\left(\tfrac{4\pi}{9}\right)\otimes Z\left(\tfrac{6\pi}{9}\right)\otimes Z\left(\tfrac{6\pi}{9}\right)\otimes Z\left(\tfrac{6\pi}{9}\right)\otimes Z\left(\tfrac{8\pi}{9}\right). \]
The corresponding weight enumerator is:
\begin{align*}
    \lambda^{*2}&=-\frac{257}{6}+\frac{4}{9}\left(6d^{2}-1\right)d\sqrt{36b^{2}+36d^{2}+2}+112b^{2}-\frac{8\sqrt{\left(162b^{4}-27b^{2}+1\right)\left(1-6d^{2}\right)}}{27\sqrt{3}}+462d^{4}\\
    &+\frac{1802d^{2}}{3}+\frac{8}{27}b\left(-\sqrt{6}\sqrt{9b^{2}-54d^{4}+27d^{2}-1}+\sqrt{\left(162b^{4}-27b^{2}+1\right)\left(18b^{2}+18d^{2}+1\right)}+6d\sqrt{6-36d^{2}}\right)
\end{align*}
\[ A=\left[1,0,\lambda^{*2},0,21-2\lambda^{*2},0,42+\lambda^{*2},0\right]+\frac{\sin^{2}(\theta)}{81}\cdot\left[0,0,0,20,-176,408,-368,116\right], \]
\[ B=\left[1,0,\lambda^{*2},21+3\lambda^{*2},21-2\lambda^{*2},126-6\lambda^{*2},42+\lambda^{*2},45+3\lambda^{*2}\right]+\frac{\sin^{2}(\theta)}{81}\cdot\left[0,0,0,-96,464,-816,624,-176\right], \]
which deviates from Equation~\eqref{eq:723shorwt-line} unless \(\sin(\theta) = 0\).

\subsection{\texorpdfstring{$\mathrm{BD}_{20}$}{BD₂₀}}

The logical basis states are:
\begin{align*}
    \sqrt{20}\left|0_{L}\right\rangle =&\left|0000000\right\rangle +i\sqrt{2}\left|0001101\right\rangle +\sqrt{3}\left|0111100\right\rangle +\left|1100110\right\rangle +is_{2}\theta\sqrt{20}\left|0010011\right\rangle +is_{3}\sqrt{4-20\theta^{2}}\left|0100011\right\rangle \\
    &+\frac{s_{4}\sqrt{7-20\theta^{2}}}{3}\left(2\left|1010101\right\rangle +\sqrt{5}\left|1011010\right\rangle \right)+\frac{s_{5}\sqrt{2+20\theta^{2}}}{3}\left(2\left|1100101\right\rangle -\sqrt{5}\left|1101010\right\rangle \right),\\
    \left|1_{L}\right\rangle =&X^{\otimes7}\left|0_{L}\right\rangle 
\end{align*}
where $s_i=\pm 1$ and $\theta\in\left[0,\sqrt{\frac{1}{5}}\right]$. The transversal logical gates are:
\[ \overline{X}=X^{\otimes7}, \]
\[ \overline{Z\left(-\tfrac{2\pi}{10}\right)}=Z\left(\tfrac{2\pi}{10}\right)\otimes Z\left(\tfrac{4\pi}{10}\right)\otimes Z\left(\tfrac{4\pi}{10}\right)\otimes Z\left(\tfrac{6\pi}{10}\right)\otimes Z\left(\tfrac{6\pi}{10}\right)\otimes Z\left(\tfrac{8\pi}{10}\right)\otimes Z\left(\tfrac{10\pi}{10}\right). \]
The weight enumerator follows Equation~\eqref{eq:723shorwt-line}, with signature norm:
\[ \lambda^{*2}=\frac{1451}{675}-\frac{1468}{135}\theta^{2}+\frac{1520}{27}\theta^{4}-\frac{16s_{2}s_{3}s_{4}s_{5}}{9}\theta\sqrt{\left(\frac{1}{5}-\theta^{2}\right)\left(\frac{7}{20}-\theta^{2}\right)\left(\frac{1}{10}+\theta^{2}\right)}. \]

\subsection{\texorpdfstring{$\mathrm{BD}_{22}$}{BD₂₂}}

The logical basis states are:
\begin{align*}
    \left|0_{L}\right\rangle =&\sqrt{\frac{5}{11}}\left(\left|0000000\right\rangle +\sqrt{2}\left|1011010\right\rangle +\sqrt{3}\left|0111100\right\rangle +2i\left|0100011\right\rangle \right)+is_{2}\theta\left|0001101\right\rangle \\
    &+is_{3}\sqrt{\frac{3}{22}-\theta^{2}}\left|0010101\right\rangle +\frac{s_{4}}{3}\sqrt{\frac{3}{11}-\theta^{2}}\left(\sqrt{5}\left|1001110\right\rangle +2\left|1101001\right\rangle \right)\\
    &+\frac{s_{5}}{3}\sqrt{\frac{3}{22}+\theta^{2}}\left(\sqrt{5}\left|1010110\right\rangle -2\left|1110001\right\rangle \right)\\
    \left|1_{L}\right\rangle =&X^{\otimes7}\left|0_{L}\right\rangle 
\end{align*}
where $s_i=\pm 1$ and $\theta\in\left[0,\sqrt{\tfrac{3}{22}}\right]$. The transversal gates are:
\[ \overline{X}=X^{\otimes7}, \]
\[ \overline{Z}\left(-\tfrac{2\pi}{11}\right)=Z\left(\tfrac{2\pi}{11}\right)\otimes Z\left(\tfrac{4\pi}{11}\right)\otimes Z\left(\tfrac{6\pi}{11}\right)\otimes Z\left(\tfrac{6\pi}{11}\right)\otimes Z\left(\tfrac{6\pi}{11}\right)\otimes Z\left(\tfrac{8\pi}{11}\right)\otimes Z\left(\tfrac{10\pi}{11}\right). \]
The weight enumerator is again given by Equation~\eqref{eq:723shorwt-line}, with signature norm:
\[ \lambda^{*2}=\frac{743}{363}-\frac{760}{99}\theta^{2}+\frac{1520}{27}\theta^{4}+\frac{16s_{2}s_{3}s_{4}s_{5}}{9}\theta\sqrt{\left(\frac{3}{11}-\theta^{2}\right)\left(\frac{3}{22}+\theta^{2}\right)\left(\frac{3}{22}-\theta^{2}\right)}. \]

\subsection{\texorpdfstring{$\mathrm{BD}_{24}$}{BD₂₄}}

The logical basis states are:
\begin{align*}
    \left|0_{L}\right\rangle =&\frac{1}{2\sqrt{2}}\left(\left|0000000\right\rangle +\left|0111010\right\rangle +\left|1011100\right\rangle -\left|1100110\right\rangle \right)\\
    &-\frac{s}{2\sqrt{3}}\left(\left|0000011\right\rangle -\left|0000101\right\rangle +\left|0111111\right\rangle -\left|1011111\right\rangle +\left|1101001\right\rangle -\left|1110001\right\rangle \right)\\
    \left|1_{L}\right\rangle =&X^{\otimes7}\left|0_{L}\right\rangle,
\end{align*}
where $s$ can take values $s=\pm 1$. In addition to the transversal logical \(X\), another transversal \(Z\)-axis rotation is implemented as:
\[ \overline{Z\left(\tfrac{2\pi}{12}\right)}=Z\left(\tfrac{2\pi}{12}\right)\otimes Z\left(\tfrac{2\pi}{12}\right)\otimes Z\left(\tfrac{6\pi}{12}\right)\otimes Z\left(\tfrac{6\pi}{12}\right)\otimes Z\left(\tfrac{10\pi}{12}\right)\otimes Z\left(\tfrac{10\pi}{12}\right)\otimes Z\left(\tfrac{14\pi}{12}\right). \]
The corresponding weight enumerator is given by Equation~\eqref{eq:723shorwt-line} with \(\lambda^* = \sqrt{\tfrac{7}{6}}\).

\subsection{\texorpdfstring{$\mathrm{BD}_{26}$}{BD₂₆}}

The logical basis states are:
\begin{align*}
    \sqrt{26}\left|0_{L}\right\rangle =&\left|0000000\right\rangle +2\left|0111010\right\rangle +\sqrt{3}\left|1011010\right\rangle +\sqrt{5}\left|1100110\right\rangle +i\left|1000011\right\rangle \\
    &+s_{2}\theta\sqrt{26}\left|1101001\right\rangle +s_{3}\sqrt{4-26\theta^{2}}\left|1110001\right\rangle +is_{4}\sqrt{6-26\theta^{2}}\left|0001101\right\rangle +is_{5}\sqrt{2+26\theta^{2}}\left|0010101\right\rangle, \\
    \left|1_{L}\right\rangle =&X^{\otimes7}\left|0_{L}\right\rangle, 
\end{align*}
where $s_i=\pm 1$ and $\theta\in\left[0,\sqrt{\tfrac{2}{13}}\right]$. The transversal logical gates are:
\[ \overline{X}=X^{\otimes7}, \]
\[ \overline{Z\left(-\tfrac{2\pi}{13}\right)}=Z\left(\tfrac{4\pi}{13}\right)\otimes Z\left(\tfrac{4\pi}{13}\right)\otimes Z\left(\tfrac{6\pi}{13}\right)\otimes Z\left(\tfrac{6\pi}{13}\right)\otimes Z\left(\tfrac{8\pi}{13}\right)\otimes Z\left(\tfrac{10\pi}{13}\right)\otimes Z\left(\tfrac{12\pi}{13}\right). \]
The Shor weight enumerator is given by Equation~\eqref{eq:723shorwt-line}, with the signature norm:
\[ \lambda^{*2}=\frac{485}{169}-\frac{160}{13}\theta^{2}+80\theta^{4}+16s_{2}s_{3}s_{4}s_{5}\theta\sqrt{\left(\frac{2}{13}-\theta^{2}\right)\left(\frac{3}{13}-\theta^{2}\right)\left(\frac{1}{13}+\theta^{2}\right)}. \]

\subsection{\texorpdfstring{$\mathrm{BD}_{28}$}{BD₂₈}}

The logical basis states are:
\begin{align*}
    \sqrt{28}\left|0_{L}\right\rangle =&\left|0000000\right\rangle +\sqrt{3}\left|0111100\right\rangle +\sqrt{5}\left|1011010\right\rangle +2\left|1110001\right\rangle +i\sqrt{6}\left|0001101\right\rangle \\
    &+s_{2}\theta\sqrt{28}\left|1010110\right\rangle +s_{3}\sqrt{5-28\theta^{2}}\left|1100110\right\rangle +is_{4}\sqrt{2-28\theta^{2}}\left|0010011\right\rangle +is_{5}\sqrt{2+28\theta^{2}}\left|0100011\right\rangle \\
    \left|1_{L}\right\rangle =&X^{\otimes7}\left|0_{L}\right\rangle 
\end{align*}
where $s_i=\pm 1$ and $\theta\in \left[0,\sqrt{\frac{1}{14}}\right]$. The transversal logical gates are:
\[ \overline{X}=X^{\otimes7}, \]
\[ \overline{Z\left(-\tfrac{2\pi}{14}\right)}=Z\left(\tfrac{4\pi}{14}\right)\otimes Z\left(\tfrac{6\pi}{14}\right)\otimes Z\left(\tfrac{6\pi}{14}\right)\otimes Z\left(\tfrac{8\pi}{14}\right)\otimes Z\left(\tfrac{8\pi}{14}\right)\otimes Z\left(\tfrac{10\pi}{14}\right)\otimes Z\left(\tfrac{12\pi}{14}\right). \]
The corresponding Shor weight enumerator is given by Equation~\eqref{eq:723shorwt-line}, with signature norm:
\[ \lambda^{*2}=\frac{95}{49}-\frac{20}{7}\theta^{2}+80\theta^{4}+16s_{2}s_{3}s_{4}s_{5}\theta\sqrt{\left(\frac{5}{28}-\theta^{2}\right)\left(\frac{1}{196}-\theta^{4}\right)}. \]

\subsection{\texorpdfstring{$\mathrm{BD}_{30}$}{BD₃₀}}

The logical basis states are:
\begin{align*}
    \sqrt{30}\left|0_{L}\right\rangle =&\left|0000000\right\rangle +\left|0000011\right\rangle +2\left|0111010\right\rangle +2\left|1011010\right\rangle +\sqrt{6}\left|1100110\right\rangle \\
    &+i\theta s_{2}\sqrt{30}\left|0001101\right\rangle +is_{3}\sqrt{9-30\theta^{2}}\left|0010101\right\rangle +s_{4}\sqrt{7-30\theta^{2}}\left|1101001\right\rangle +s_{5}\sqrt{30\theta^{2}-2}\left|1110001\right\rangle \\
    \left|1_{L}\right\rangle =&X^{\otimes7}\left|0_{L}\right\rangle 
\end{align*}
where $s_i=\pm 1$ and $\theta\in \left[\sqrt{\frac{1}{15}},\sqrt{\frac{7}{30}}\right]$. The transversal logical gates are:
\[ \overline{X}=X^{\otimes7}, \]
\[ \overline{Z\left(-\tfrac{2\pi}{15}\right)}=Z\left(\tfrac{4\pi}{15}\right)\otimes Z\left(\tfrac{4\pi}{15}\right)\otimes Z\left(\tfrac{6\pi}{15}\right)\otimes Z\left(\tfrac{6\pi}{15}\right)\otimes Z\left(\tfrac{8\pi}{15}\right)\otimes Z\left(\tfrac{14\pi}{15}\right)\otimes Z\left(\tfrac{16\pi}{15}\right). \]
The Shor weight enumerator is given by Equation~\eqref{eq:723shorwt-line}, with signature norm:
\[ \lambda^{*2}=\frac{313}{75}-24\theta^{2}+80\theta^{4}+16s_{2}s_{3}s_{4}s_{5}\theta\sqrt{\left(\frac{3}{10}-\theta^{2}\right)\left(\frac{7}{30}-\theta^{2}\right)\left(\theta^{2}-\frac{1}{15}\right)}. \]

\subsection{\texorpdfstring{$\mathrm{BD}_{34}$}{BD₃₄}}

The logical basis states are:
\begin{align*}
    \sqrt{34}\left|0_{L}\right\rangle =&e^{i\theta}\left|0000000\right\rangle +s_{2}e^{i\theta}\left|0000011\right\rangle +2s_{3}\left|0111010\right\rangle +2s_{4}\left|1011010\right\rangle \\
    &+\sqrt{6}s_{5}\left|1100110\right\rangle +\sqrt{7}s_{6}\left|1101001\right\rangle +i\sqrt{2}s_{7}\left|0001110\right\rangle +3i\left|0010101\right\rangle ,\\
    \left|1_{L}\right\rangle =&X^{\otimes7}\left|0_{L}\right\rangle ,
\end{align*}
where \(s_i = \pm 1\) and \(\theta \in \mathbb{R}\). The transversal group \(\mathrm{BD}_{34}\) is generated by the following two logical gates:
\[ \overline{X}=X^{\otimes7}, \]
\[ \overline{Z\left(-\tfrac{2\pi}{17}\right)}=Z\left(\tfrac{4\pi}{17}\right)\otimes Z\left(\tfrac{4\pi}{17}\right)\otimes Z\left(\tfrac{6\pi}{17}\right)\otimes Z\left(\tfrac{8\pi}{17}\right)\otimes Z\left(\tfrac{10\pi}{17}\right)\otimes Z\left(\tfrac{16\pi}{17}\right)\otimes Z\left(\tfrac{18\pi}{17}\right).\]
The quantum weight enumerator for this code is:
\[ A=\left[1,0,\frac{831}{289},\frac{44}{289},\frac{4063}{289},\frac{768}{289},\frac{12289}{289},\frac{212}{289}\right]+\cos(2\theta)\cdot\left[0,0,0,-\frac{44}{289},\frac{344}{289},-\frac{768}{289},\frac{680}{289},-\frac{212}{289}\right], \]
\[ B=\left[1,0,\frac{831}{289},\frac{8394}{289},\frac{5255}{289},\frac{29892}{289},\frac{14169}{289},\frac{15154}{289}\right]+\cos(2\theta)\cdot\left[0,0,0,\frac{168}{289},-\frac{848}{289},\frac{1536}{289},-\frac{1200}{289},\frac{344}{289}\right]. \]

\section{More \texorpdfstring{$((8,2,3))$}{((8,2,3))} Codes with Transversal Group \texorpdfstring{$\mathrm{BD}_{2m}$}{BD₂ₘ}}\label{appendix:823-BD2m}

\subsection{\texorpdfstring{$\mathrm{BD}_{38}$}{BD₃₈}}

The logical states are:
\begin{align*}
    \left|0_{L}\right\rangle =&\left(\theta_{1}+i\theta_{2}\right)\left|00000000\right\rangle +i\theta_{3}\left|00100011\right\rangle +i\theta_{4}\left|00101110\right\rangle +i\theta_{5}\left|00110110\right\rangle +i\theta_{6}\left|01000011\right\rangle \\&+i\theta_{7}\left|01001110\right\rangle +i\theta_{8}\left|01010110\right\rangle +\theta_{9}\left|01101001\right\rangle +\theta_{10}\left|01110001\right\rangle +\theta_{11}\left|01111100\right\rangle \\&+\theta_{12}\left|10011001\right\rangle +\theta_{13}\left|10100101\right\rangle +\theta_{14}\left|11000101\right\rangle +\theta_{15}\left|11101010\right\rangle +\theta_{16}\left|11110010\right\rangle \\
    \left|1_{L}\right\rangle =&X^{\otimes8}\left|0_{L}\right\rangle 
\end{align*}
while $\theta_i^2$ satisfying the following linear identities:
\begingroup
\setcounter{MaxMatrixCols}{20} 
\[ C=\begin{bmatrix}
1 & 1 & 1 & 1 & 1 & 1 & 1 & 1 & 1 & 1 & 1 & -1 & -1 & -1 & -1 & -1\\
1 & 1 & 1 & 1 & 1 & -1 & -1 & -1 & -1 & -1 & -1 & 1 & 1 & -1 & -1 & -1\\
1 & 1 & -1 & -1 & -1 & 1 & 1 & 1 & -1 & -1 & -1 & 1 & -1 & 1 & -1 & -1\\
1 & 1 & 1 & 1 & -1 & 1 & 1 & -1 & 1 & -1 & -1 & -1 & 1 & 1 & 1 & -1\\
1 & 1 & 1 & -1 & 1 & 1 & -1 & 1 & -1 & 1 & -1 & -1 & 1 & 1 & -1 & 1\\
1 & 1 & 1 & -1 & -1 & 1 & -1 & -1 & 1 & 1 & -1 & 1 & -1 & -1 & 1 & 1\\
1 & 1 & -1 & -1 & -1 & -1 & -1 & -1 & 1 & 1 & 1 & 1 & 1 & 1 & -1 & -1\\
1 & 1 & -1 & 1 & 1 & -1 & 1 & 1 & -1 & -1 & 1 & -1 & -1 & -1 & 1 & 1\\
\end{bmatrix}, \]
\endgroup
\[ \begin{cases}
C\cdot\left[\theta_{1}^{2},\theta_{2}^{2},\cdots,\theta_{16}^{2}\right]=0\\
\sum_{i}\theta_{i}^{2}=1
\end{cases}.\]
From these constraints, some variables can be solved $\theta_1^2+\theta_2^2=\frac{1}{38}$ and $\theta_{12}^2=\frac{4}{19}$. Above are $16$ real variables and $9$ linear independent linear constraints, thus in general there will be $7$ degree of freedom. Transversal gates are:
\[ \overline{X}=X^{\otimes8}, \]
\[ \overline{Z\left(-\tfrac{2\pi}{19}\right)}=Z\left(\tfrac{4\pi}{19}\right)\otimes Z\left(\tfrac{6\pi}{19}\right)\otimes Z\left(\tfrac{6\pi}{19}\right)\otimes Z\left(\tfrac{8\pi}{19}\right)\otimes Z\left(\tfrac{8\pi}{19}\right)\otimes Z\left(\tfrac{10\pi}{19}\right)\otimes Z\left(\tfrac{14\pi}{19}\right)\otimes Z\left(\tfrac{18\pi}{19}\right).\]

\subsection{\texorpdfstring{$\mathrm{BD}_{72}$}{BD₇₂}}

An example of $((8,2,3))$ code with transversal group $\mathrm{BD}_{72}$:
\begin{align*}
    \sqrt{72}\left|0_{L}\right\rangle =&\left|00000000\right\rangle +\sqrt{14}e^{i\theta_{1}}\left|00001101\right\rangle +\sqrt{3}e^{i\theta_{2}}\left|00111100\right\rangle +\sqrt{13}s_{1}e^{i\theta_{2}}\left|01011010\right\rangle +\sqrt{5}e^{i\theta_{3}}\left|01100110\right\rangle \\&+\sqrt{8}e^{i\theta_{4}}\left|10010110\right\rangle +\sqrt{10}is_{2}e^{i\theta_{2}}\left|10100011\right\rangle +\sqrt{6}e^{i\theta_{5}}\left|11101100\right\rangle +\sqrt{12}is_{3}e^{i\theta_{1}}\left|11110001\right\rangle, 
\end{align*}
where $s_i=\pm 1$ and $\theta_i\in\mathbb{R}$. Logical 1 is obtained via transversal gate $\overline{X}=X^{\otimes 8}$ and another transversal Z-axis rotation is implemented as:
\[ \overline{Z\left(-\tfrac{2\pi}{36}\right)}=Z\left(\tfrac{6\pi}{36}\right)\otimes Z\left(\tfrac{10\pi}{36}\right)\otimes Z\left(\tfrac{12\pi}{36}\right)\otimes Z\left(\tfrac{16\pi}{36}\right)\otimes Z\left(\tfrac{20\pi}{36}\right)\otimes Z\left(\tfrac{24\pi}{36}\right)\otimes Z\left(\tfrac{26\pi}{36}\right)\otimes Z\left(\tfrac{28\pi}{36}\right). \]

\subsection{\texorpdfstring{$\mathrm{BD}_{74}$}{BD₇₄}}

An example of $((8,2,3))$ code with transversal group $\mathrm{BD}_{74}$:
\begin{align*}
    \sqrt{74}\left|0_{L}\right\rangle =&\left|00000000\right\rangle +\sqrt{12}e^{i\theta_{1}}\left|00000111\right\rangle +\sqrt{3}e^{i\theta_{2}}\left|00111100\right\rangle +\sqrt{11}s_{1}e^{i\theta_{2}}\left|01011010\right\rangle +\sqrt{10}e^{i\theta_{3}}\left|01101001\right\rangle \\&+\sqrt{6}is_{2}e^{i\theta_{3}}\left|10001110\right\rangle +\sqrt{7}e^{i\theta_{4}}\left|10011001\right\rangle +\sqrt{8}is_{3}e^{i\theta_{2}}\left|10100011\right\rangle +\sqrt{16}is_{4}e^{i\theta_{1}}\left|11110100\right\rangle 
\end{align*}
where $s_i=\pm 1$ and $\theta_i\in\mathbb{R}$. Logical 1 is obtained via transversal gate $\overline{X}=X^{\otimes 8}$ and another transversal Z-axis rotation is implemented as:
\[ \overline{Z\left(-\tfrac{2\pi}{37}\right)}=Z\left(\tfrac{8\pi}{37}\right)\otimes Z\left(\tfrac{12\pi}{37}\right)\otimes Z\left(\tfrac{14\pi}{37}\right)\otimes Z\left(\tfrac{18\pi}{37}\right)\otimes Z\left(\tfrac{20\pi}{37}\right)\otimes Z\left(\tfrac{22\pi}{37}\right)\otimes Z\left(\tfrac{24\pi}{37}\right)\otimes Z\left(\tfrac{28\pi}{37}\right). \]

\subsection{\texorpdfstring{$\mathrm{BD}_{76}$}{BD₇₆}}

An example of $((8,2,3))$ code with transversal group $\mathrm{BD}_{76}$:
\begin{align*}
    \sqrt{76}\left|0_{L}\right\rangle =&\left|00000000\right\rangle +\sqrt{18}e^{i\theta_{1}}\left|00001101\right\rangle +2e^{i\theta_{2}}\left|00100011\right\rangle +\sqrt{3}e^{i\theta_{3}}\left|00111100\right\rangle +\sqrt{7}e^{i\theta_{4}}\left|01011010\right\rangle \\&+\sqrt{5}e^{i\theta_{5}}\left|01100110\right\rangle +\sqrt{12}e^{i\theta_{6}}\left|10010110\right\rangle +\sqrt{10}is_{1}e^{i\theta_{1}}\left|11101010\right\rangle +4is_{2}e^{i\theta_{1}}\left|11110001\right\rangle 
\end{align*}
where $s_i=\pm 1$ and $\theta_i\in\mathbb{R}$. Logical 1 is obtained via transversal gate $\overline{X}=X^{\otimes 8}$ and another transversal Z-axis rotation is implemented as:
\[ \overline{Z\left(-\tfrac{2\pi}{38}\right)}=Z\left(\tfrac{4\pi}{38}\right)\otimes Z\left(\tfrac{8\pi}{38}\right)\otimes Z\left(\tfrac{14\pi}{38}\right)\otimes Z\left(\tfrac{18\pi}{38}\right)\otimes Z\left(\tfrac{20\pi}{38}\right)\otimes Z\left(\tfrac{24\pi}{38}\right)\otimes Z\left(\tfrac{30\pi}{38}\right)\otimes Z\left(\tfrac{32\pi}{38}\right).\]

\subsection{\texorpdfstring{$\mathrm{BD}_{78}$}{BD₇₈}}

An example of $((8,2,3))$ code with transversal group $\mathrm{BD}_{78}$:
\begin{align*}
    \sqrt{78}\left|0_{L}\right\rangle =&\left|00000000\right\rangle +4e^{i\theta_{1}}\left|00010011\right\rangle +\sqrt{3}e^{i\theta_{2}}\left|00111100\right\rangle +\sqrt{5}e^{i\theta_{3}}\left|01100110\right\rangle +\sqrt{14}e^{i\theta_{4}}\left|01101001\right\rangle \\&+\sqrt{10}is_{1}e^{i\theta_{4}}\left|10001110\right\rangle +3e^{i\theta_{5}}\left|10100101\right\rangle +\sqrt{12}is_{1}e^{i\theta_{1}}\left|11011100\right\rangle +\sqrt{8}e^{i\theta_{6}}\left|11110010\right\rangle 
\end{align*}
where $s_i=\pm 1$ and $\theta_i\in\mathbb{R}$. Logical 1 is obtained via transversal gate $\overline{X}=X^{\otimes 8}$ and another transversal Z-axis rotation is implemented as:
\[ \overline{Z\left(-\tfrac{2\pi}{39}\right)}=Z\left(\tfrac{6\pi}{39}\right)\otimes Z\left(\tfrac{10\pi}{39}\right)\otimes Z\left(\tfrac{16\pi}{39}\right)\otimes Z\left(\tfrac{18\pi}{39}\right)\otimes Z\left(\tfrac{20\pi}{39}\right)\otimes Z\left(\tfrac{24\pi}{39}\right)\otimes Z\left(\tfrac{28\pi}{39}\right)\otimes Z\left(\tfrac{32\pi}{39}\right). \]

\subsection{\texorpdfstring{$\mathrm{BD}_{80}$}{BD₈₀}}

An example of $((8,2,3))$ code with transversal group $\mathrm{BD}_{80}$:
\begin{align*}
    \sqrt{80}\left|0_{L}\right\rangle =&\left|00000000\right\rangle +\sqrt{2}e^{i\theta_{1}}\left|00001110\right\rangle +\sqrt{8}e^{i\theta_{2}}\left|00010101\right\rangle +\sqrt{18}e^{i\theta_{3}}\left|00111010\right\rangle +\sqrt{5}e^{i\theta_{4}}\left|01100110\right\rangle \\&+\sqrt{6}e^{i\theta_{5}}\left|01101001\right\rangle +\sqrt{11}is_{1}e^{i\theta_{3}}\left|10100101\right\rangle +\sqrt{15}is_{2}e^{i\theta_{3}}\left|11000011\right\rangle +\sqrt{14}e^{i\theta_{6}}\left|11011100\right\rangle ,\\
    \left|1_{L}\right\rangle =&X^{\otimes8}\left|0_{L}\right\rangle.
\end{align*}
where \(s_i = \pm 1\) and \(\theta_i \in \mathbb{R}\) are free parameters. Logical \(\ket{1_L}\) is obtained by applying the transversal operator \(\overline{X} = X^{\otimes 8}\) to \(\ket{0_L}\). An additional transversal \(Z\)-axis rotation is implemented as:
\[ \overline{Z\left(-\tfrac{2\pi}{40}\right)}=Z\left(\tfrac{4\pi}{40}\right)\otimes Z\left(\tfrac{10\pi}{40}\right)\otimes Z\left(\tfrac{12\pi}{40}\right)\otimes Z\left(\tfrac{16\pi}{40}\right)\otimes Z\left(\tfrac{22\pi}{40}\right)\otimes Z\left(\tfrac{28\pi}{40}\right)\otimes Z\left(\tfrac{30\pi}{40}\right)\otimes Z\left(\tfrac{36\pi}{40}\right).\]

\bibliographystyle{plain}
\bibliography{reference}

\begin{thebibliography}{10}

\bibitem{aggarwal2008boolean}
Vaneet Aggarwal and A~Robert Calderbank.
\newblock Boolean functions, projection operators, and quantum error correcting codes.
\newblock {\em IEEE Transactions on Information Theory}, 54(4):1700--1707, 2008.

\bibitem{anderson2016classification}
Jonas~T Anderson and Tomas Jochym-O'Connor.
\newblock Classification of transversal gates in qubit stabilizer codes.
\newblock {\em Quantum Information \& Computation}, 16(9-10):771--802, 2016.

\bibitem{arvind2002nonstabilizer}
Vikraman Arvind, Piyush~P Kurur, and KR~Parthasarathy.
\newblock Nonstabilizer quantum codes from abelian subgroups of the error group.
\newblock {\em arXiv preprint quant-ph/0210097}, 2002.

\bibitem{calderbank1998quantum}
A.R. Calderbank, E.M. Rains, P.M. Shor, and N.J.A. Sloane.
\newblock Quantum error correction via codes over gf(4).
\newblock {\em IEEE Transactions on Information Theory}, 44(4):1369--1387, 1998.

\bibitem{Cao_2022}
Chenfeng Cao, Chao Zhang, Zipeng Wu, Markus Grassl, and Bei Zeng.
\newblock Quantum variational learning for quantum error-correcting codes.
\newblock {\em Quantum}, 6:828, October 2022.

\bibitem{chuang2009codeword}
Isaac Chuang, Andrew Cross, Graeme Smith, John Smolin, and Bei Zeng.
\newblock Codeword stabilized quantum codes: Algorithm and structure.
\newblock {\em Journal of Mathematical Physics}, 50(4), 2009.

\bibitem{cross2009codeword}
Andrew Cross, Graeme Smith, John~A Smolin, and Bei Zeng.
\newblock Codeword stabilized quantum codes.
\newblock {\em IEEE transactions on information theory}, 55(1):433--438, 2009.

\bibitem{cross2025small}
Andrew Cross and Drew Vandeth.
\newblock Small binary stabilizer subsystem codes.
\newblock {\em arXiv preprint arXiv:2501.17447}, 2025.

\bibitem{PhysRev.93.99}
R.~H. Dicke.
\newblock Coherence in spontaneous radiation processes.
\newblock {\em Phys. Rev.}, 93:99--110, Jan 1954.

\bibitem{dixon1970computing}
John~D Dixon.
\newblock Computing irreducible representations of groups.
\newblock {\em Mathematics of Computation}, 24(111):707--712, 1970.

\bibitem{du2024characterizing}
Mengxin Du, Chao Zhang, Yiu-Tung Poon, and Bei Zeng.
\newblock Characterizing quantum codes via the coefficients in knill-laflamme conditions.
\newblock {\em arXiv preprint arXiv:2410.07983}, 2024.

\bibitem{eastin2009restrictions}
Bryan Eastin and Emanuel Knill.
\newblock Restrictions on transversal encoded quantum gate sets.
\newblock {\em Physical review letters}, 102(11):110502, 2009.

\bibitem{gottesman1997stabilizer}
Daniel Gottesman.
\newblock Stabilizer codes and quantum error correction.
\newblock {\em arXiv: Quantum Physics}, 1997.

\bibitem{grassl1997note}
Markus Grassl and Thomas Beth.
\newblock A note on non-additive quantum codes.
\newblock {\em arXiv preprint quant-ph/9703016}, 1997.

\bibitem{grassl2008non}
Markus Grassl and Martin Rotteler.
\newblock Non-additive quantum codes from goethals and preparata codes.
\newblock In {\em 2008 IEEE Information Theory Workshop}, pages 396--400. IEEE, 2008.

\bibitem{knill1997theory}
Emanuel Knill and Raymond Laflamme.
\newblock Theory of quantum error-correcting codes.
\newblock {\em Phys. Rev. A}, 55:900--911, Feb 1997.

\bibitem{koutsioumpas2024quantum}
Stergios Koutsioumpas.
\newblock {\em Quantum Error Correction with Transversal non-Clifford Gates}.
\newblock PhD thesis, Royal Holloway University of London, 2024.

\bibitem{kubischta2023family}
Eric Kubischta and Ian Teixeira.
\newblock Family of quantum codes with exotic transversal gates.
\newblock {\em Physical Review Letters}, 131(24):240601, 2023.

\bibitem{kubischta2024permutationinvariantquantumcodestransversal}
Eric Kubischta and Ian Teixeira.
\newblock Permutation-invariant quantum codes with transversal generalized phase gates, 2024.

\bibitem{lezcano2019trivializations}
Mario Lezcano~Casado.
\newblock Trivializations for gradient-based optimization on manifolds.
\newblock {\em Advances in Neural Information Processing Systems}, 32, 2019.

\bibitem{lidar1998decoherence}
D.~A. Lidar, I.~L. Chuang, and K.~B. Whaley.
\newblock Decoherence-free subspaces for quantum computation.
\newblock {\em Phys. Rev. Lett.}, 81:2594--2597, Sep 1998.

\bibitem{lidar2003decoherence}
Daniel~A. Lidar and K.~Birgitta~Whaley.
\newblock Decoherence-free subspaces and subsystems.
\newblock In Fabio Benatti and Roberto Floreanini, editors, {\em Irreversible Quantum Dynamics}, pages 83--120. Springer Berlin Heidelberg, Berlin, Heidelberg, 2003.

\bibitem{miller2024experimental}
Daniel Miller, Kyano Levi, Lukas Postler, Alex Steiner, Lennart Bittel, Gregory~AL White, Yifan Tang, Eric~J Kuehnke, Antonio~A Mele, Sumeet Khatri, et~al.
\newblock Experimental measurement and a physical interpretation of quantum shadow enumerators.
\newblock {\em arXiv preprint arXiv:2408.16914}, 2024.

\bibitem{ni2015non}
Xiaotong Ni, Oliver Buerschaper, and Maarten Van~den Nest.
\newblock A non-commuting stabilizer formalism.
\newblock {\em Journal of Mathematical Physics}, 56(5), 2015.

\bibitem{ouyang2014permutation}
Yingkai Ouyang.
\newblock Permutation-invariant quantum codes.
\newblock {\em Physical Review A}, 90(6):062317, 2014.

\bibitem{ouyang2021permutation}
Yingkai Ouyang.
\newblock Permutation-invariant quantum coding for quantum deletion channels.
\newblock In {\em 2021 IEEE International Symposium on Information Theory (ISIT)}, pages 1499--1503. IEEE, 2021.

\bibitem{pollatsek2004permutationally}
Harriet Pollatsek and Mary~Beth Ruskai.
\newblock Permutationally invariant codes for quantum error correction.
\newblock {\em Linear Algebra and its Applications}, 392:255--288, 2004.

\bibitem{rains1998quantum}
E.M. Rains.
\newblock Quantum weight enumerators.
\newblock {\em IEEE Transactions on Information Theory}, 44(4):1388--1394, 1998.

\bibitem{rains1999quantum}
E.M. Rains.
\newblock Quantum shadow enumerators.
\newblock {\em IEEE Transactions on Information Theory}, 45(7):2361--2366, 1999.

\bibitem{rains1999quantumCodes}
Eric~M. Rains.
\newblock Quantum codes of minimum distance two.
\newblock {\em IEEE Transactions on Information theory}, 45(1):266--271, 1999.

\bibitem{rains1997nonadditive}
Eric~M. Rains, R.~H. Hardin, Peter~W. Shor, and N.~J.~A. Sloane.
\newblock A nonadditive quantum code.
\newblock {\em Phys. Rev. Lett.}, 79:953--954, Aug 1997.

\bibitem{roychowdhury1997structure}
Vwani~P Roychowdhury and Farrokh Vatan.
\newblock On the structure of additive quantum codes and the existence of nonadditive codes.
\newblock {\em arXiv preprint quant-ph/9710031}, 1997.

\bibitem{shor1997quantum}
Peter Shor and Raymond Laflamme.
\newblock Quantum analog of the macwilliams identities for classical coding theory.
\newblock {\em Phys. Rev. Lett.}, 78:1600--1602, Feb 1997.

\bibitem{steane1996error}
A.~M. Steane.
\newblock Error correcting codes in quantum theory.
\newblock {\em Phys. Rev. Lett.}, 77:793--797, Jul 1996.

\bibitem{steane1996simple}
A.~M. Steane.
\newblock Simple quantum error-correcting codes.
\newblock {\em Phys. Rev. A}, 54:4741--4751, Dec 1996.

\bibitem{webster2022xp}
Mark~A Webster, Benjamin~J Brown, and Stephen~D Bartlett.
\newblock The xp stabiliser formalism: a generalisation of the pauli stabiliser formalism with arbitrary phases.
\newblock {\em Quantum}, 6:815, 2022.

\bibitem{yu2008nonadditive}
Sixia Yu, Qing Chen, CH~Lai, and CH~Oh.
\newblock Nonadditive quantum error-correcting code.
\newblock {\em Physical review letters}, 101(9):090501, 2008.

\bibitem{yu2007graphical}
Sixia Yu, Qing Chen, and Choo~Hiap Oh.
\newblock Graphical quantum error-correcting codes.
\newblock {\em arXiv preprint arXiv:0709.1780}, 2007.

\bibitem{zeng2011transversality}
Bei Zeng, Andrew Cross, and Isaac~L Chuang.
\newblock Transversality versus universality for additive quantum codes.
\newblock {\em IEEE Transactions on Information Theory}, 57(9):6272--6284, 2011.

\end{thebibliography}

\end{document}